\documentclass[reprint,prx,aps,amsmath,amssymb,floatfix,
superscriptaddress,longbibliography]{revtex4-1}

\usepackage[T1]{fontenc}
\usepackage{dcolumn}
\usepackage{bm}
\usepackage{graphicx}
\usepackage{amsthm}
\usepackage{color}
\usepackage{xcolor}
\usepackage{verbatim}
\usepackage{esint}
\usepackage[english]{babel}
\usepackage{amsfonts}
\usepackage{slashed}
\usepackage{latexsym}
\usepackage{bm}
\usepackage[colorlinks = true,
            linkcolor = red,
            urlcolor  = blue,
            citecolor = magenta,
            anchorcolor = red]{hyperref}
\usepackage{esint}
\usepackage{soul}
\usepackage{cancel}
\usepackage[normalem]{ulem}
\usepackage{empheq}
\usepackage{dsfont}
\usepackage{amsmath}
\usepackage{epsfig} 
\usepackage{enumerate}
\definecolor{sapphire}{rgb}{0.03, 0.03, 0.41}
\DeclareMathOperator{\sech}{sech}

\DeclareMathAlphabet\mathbfcal{OMS}{cmsy}{b}{n}
\def\be{\begin{equation}}
\def\ee{\end{equation}}

\begin{document}

\title{Geometric properties of adiabatic quantum thermal machines}

\author{Bibek Bhandari}
\affiliation{NEST, Scuola Normale Superiore and Istituto Nanoscienze-CNR, I-56126 Pisa, Italy}

\author{Pablo Terr\'en Alonso}
\affiliation{International Center for Advanced Studies, Escuela de Ciencia y Tecnolog\'ia, Universidad Nacional de San Mart\'in, 
Avenida 25 de Mayo y Francia, 1650 Buenos Aires, Argentina\:}

\author{Fabio Taddei}
\affiliation{NEST, Istituto Nanoscienze-CNR and Scuola Normale Superiore, I-56126 Pisa, Italy\:}

\author{Felix von Oppen}
\affiliation{Dahlem Center for Complex Quantum Systems and Fachbereich Physik, Freie Universit\"at Berlin, Arnimallee 14, 14195 Berlin, Germany}

\author{Rosario Fazio}
\affiliation{Abdus Salam ICTP, Strada Costiera 11, I-34151 Trieste, Italy}
\affiliation{Dipartimento di Fisica, Universit\`a di Napoli ``Federico II'', Monte S. Angelo, I-80126 Napoli, Italy}

\author{Liliana Arrachea}
\affiliation{International Center for Advanced Studies, Escuela de Ciencia y Tecnolog\'ia, Universidad Nacional de San Mart\'in, 
Avenida 25 de Mayo y Francia, 1650 Buenos Aires, Argentina\:}

\date{\today}

\begin{abstract}
We present a general unified approach for the study of quantum thermal machines, including both heat engines and refrigerators, operating under periodic adiabatic driving and in contact with thermal reservoirs kept at different temperatures. We show that many observables  characterizing this operating mode and the performance of the machine are of geometric nature. Heat-work conversion mechanisms and dissipation of energy can be described, respectively, by the antisymmetric and symmetric components of a thermal geometric tensor defined in the space of time-dependent parameters generalized to include the temperature bias. The antisymmetric component can be identified as a Berry curvature, while the symmetric component defines the metric of the manifold. We show that the operation of adiabatic thermal machines, and consequently also their efficiency, are intimately related to these geometric aspects. We illustrate these ideas by discussing 
two specific cases: a slowly driven qubit  asymmetrically coupled to two bosonic reservoirs kept at different temperatures, and a 
quantum dot driven by a rotating magnetic field and strongly coupled to electron reservoirs with different polarizations. Both examples are already amenable for an experimental verification.
\end{abstract}

\maketitle
\section{Introduction}
Thermodynamics in quantum nanoscale systems~\cite{scovil1959,geusic1967,geva1992,allahverdyan2000,hanggi2009,gemmer2009,
horodecki2013,binder2019} has been a rapidly growing research topic for some years now, emerging at the intersection of statistical 
mechanics, nanoscience, quantum information, as well as atomic and molecular physics. A paradigmatic goal in this field is to conceive 
of and realize thermal machines in the quantum realm, which, like the classical thermodynamic cycles, transform heat to useful work or 
use work to refrigerate~\cite{alicki1979,geva1992b,linden2010,cycle1,cycle2,cycle3,cycle4,cycle5,cycle6,cycle7,cycle8,cycle9,cycle10}. 
The development of efficient thermal machines operating in the quantum realm is, in fact, of paramount importance also for 
quantum technologies. Numerous theoretical proposals~\cite{pekola2007,murphy2008,esposito2009,janine1,kurizki2013,sothmann2014,
kosloff2014,hofer2016,benenti2017,nanomech2,nanomech3} stimulated experimental efforts on several platforms~\cite{peterson2019,
roche,pothier}, including solid-state electronics~\cite{electro1,molenkamp2015,pekola2019,ronzani} and nanomechanical systems~\cite{nanomech0,
nanomech1,nanomech4,nanomech5,nanomech6}, as well as cold atoms and trapped ions~\cite{coldat,expotto,superad,rossnagel2016,assis2019}.

  \begin{figure}[htb]
	\centering
	\includegraphics[width=\columnwidth]{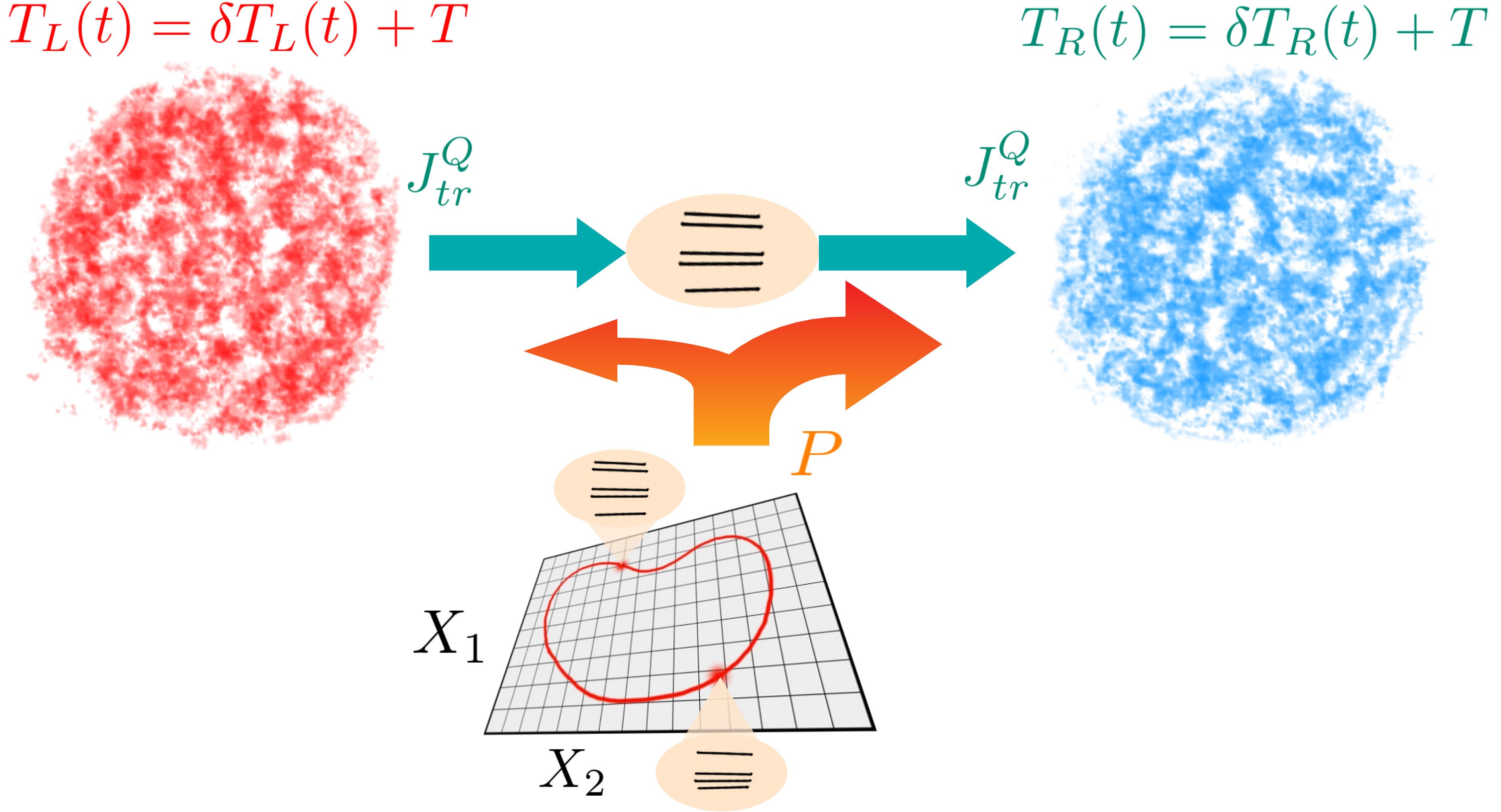}
\caption{Geometrical thermal machine setup. A central, parametrically driven quantum system described by the Hamiltonian ${\cal H_S}$ is coupled to macroscopic reservoirs. A cycle of the machine is completely characterized by a closed path in the parameter space ${\bf X}$. After a complete cycle the averaged power $P$ is dissipated as heat  in the reservoirs. The net transported energy $J^Q_{\rm tr}$ flows from one reservoir to the other.}
	\label{generic-q-machine}
\end{figure}

In its most simplified version a quantum thermal machine is composed of a working substance (typically a few-level quantum system) coupled 
to two or more thermal baths kept at different temperatures (and possibly at different chemical potentials). Engines and refrigerators can operate 
under steady-state conditions, as thermoelectric engines, or be controlled by time-periodic perturbations which define a cycle, as in conventional 
macroscopic thermal machines. An example of the latter is the quantum Otto engine, which has been investigated 
theoretically \cite{kosloff84,geva1992b,kosloff00,scully02,lin03,kieu04,Rezek2006,Quan2007,henrich2007,he2009,abah12,Campo:2014aa,alecce2015,
legggio2016,karimi,Kosloff2017,Watanabe2017,Campisi2016,camp,cycle4,jongmin2019}  and realized experimentally~\cite{expotto,assis2019,peterson2019}. 
Understanding how to discriminate and characterize useful work, heat, and dissipated energy in these systems is a fundamental step towards 
the realization of nanomachines. In fact, unlike the ideal classical thermodynamic cycles, quantum  thermal machines typically operate out of
equilibrium~\cite{gallego2014,anders2016}, which necessarily implies entropy production and dissipation. In addition to its impact on emerging 
technologies, the study of quantum heat engines and refrigerators is also of fundamental importance to deepen our understanding of how energy 
flows and transforms at the nanoscale  \cite{kieu2004,uzdin2015,seifert,benenti2017,palao2019,binder,barra}.

In the present work we will consider adiabatically driven thermal machines. Their cycle is controlled by time-periodic  
 changes of a set of parameters which are slow compared to the typical time scales associated with the (quantum) working 
 substance (see for example Refs.~\onlinecite{cavina,brandner}).  The modulation can be associated with parameters of the baths (temperature, chemical 
 potential, $\dots$ ) or the working medium (external fields, coupling constants, $\dots$), see Fig.\ \ref{generic-q-machine}. 
  We will refer to these quantum machines as {\em geometric thermal machines}. In this regime and for small amplitude of the thermal bias, the operation 
  has a purely geometric description. At the heart of this description is  the thermal geometric tensor introduced in Section~\ref{geoF}.
 {\em  Within the  adiabatic linear response regime, the process of heat-work conversion is related to the antisymmetric component of the  thermal geometric tensor, while the dissipation and entropy production are related to the 
 symmetric component of the same tensor.} Importantly, the antisymmetric component
 has the structure of a Berry curvature, which depends only on the geometry 
 of the cycle in parameter space,
 and can be straightforwardly expressed in terms of a line integral in this space. This representation is very useful for identifying optimal protocols of heat-work conversion. Furthermore, the symmetric component has a geometric interpretation in terms of thermodynamic length and can also be represented as a line integral for cyclic protocols, which is useful in the design of efficient 
protocols. 
Our approach does not only allow one to describe a whole class of quantum machines in a unifying picture. It also has 
practical implications such as improved ways to optimize their performance, as we illustrate by two paradigmatic systems: a qubit and a quantum dot.

Starting from the seminal works of Aharonov and Bohm~\cite{ab} as well as Berry~\cite{berry}, geometric effects have pervaded many areas of physics. In quantum transport, distinct contributions of geometric origin affect charge and energy currents. In the absence of an additional {\em dc} bias, the pumped charge in a periodically driven system was shown to be of geometric origin, and can thus be expressed in terms of a closed-path integral in parameter space~\cite{brouwer,aleiner,avron,janine,
hasegawa2017,hasegawa2018},  akin to the Berry phase~\cite{berry}.  A similar approach was adopted to analyze heat transport in a driven two-level system weakly coupled to bosonic baths \cite{hanggi}. Closely 
related to these ideas is the geometric description of driving-induced forces~\cite{berry2010,miche,Bode2011,Bode2012,Thomas2012,raul1,raul2,raul3,pablo}, including geometric magnetism~\cite{Berry1993,campisi2}, with the extension of geometric response functions to open systems also being discussed in relation to Cooper pair pumping~\cite{campisi3}. 
Geometric concepts like a {\em thermodynamic metric} and a {\em thermodynamic length} were recently introduced as promising tools to characterize the dissipated energy and to design optimal driving protocols \cite{reza,crooks,deff,marti,marti2,marti3}. 
Similar ideas are behind the description of the
 adiabatic time-evolution of many-body ground states of closed systems  in terms of a {\em geometric tensor} \cite{anatoli1,anatoli2,anatoli3}. The topological characterization of mixed thermal states is also close to these  concepts\cite{diehl1,diehl2}.

This large body of work linking geometry to transport naturally hints at similar connections for thermal machines. First, thermal machines involve periodic variations of parameters and one may naturally expect geometric effects in the sense of Berry to play an important role. Second, the efficiency with which thermal machines operate is reduced by dissipation, and thus geometry enters the physics of thermal machines also in a second rather distinct way through the concept of thermodynamic length. In the present paper, under quite general assumptions, we will show that the operation of quantum thermal machines and the underlying heat-work conversion is fundamentally tied to such geometric effects. We formulate a unified description in terms of a geometric tensor for all the relevant energy fluxes, which we refer to as {\em thermal geometric tensor}.
Within this description, pumping and dissipation are, respectively, associated with the antisymmetric and symmetric components of this tensor. We also show that not only heat pumping but also the dissipated heat can be characterized in terms of  an integral over a closed path in parameter space. These results apply universally to any periodically and adiabatically driven quantum system in contact with various reservoirs, irrespective of the statistics obeyed by the particles, the strength of the coupling between the system and the reservoir, or the 
presence of many-body interactions.  
  
The article is organized as follows.  In Section\ \ref{model-heat-engine}, we introduce the model of an adiabatic thermal machine. We also introduce the linear-response formalism to treat {\em ac} adiabatic and thermal driving. Section~\ref{geoF} is devoted to the analysis of the thermodynamic behavior of the heat engine. This section contains the principal results of the present work and shows how the performance characteristics of the engine (efficiency, output power, etc.) are of geometric origin. The central results of this approach are captured by Eqs.~(\ref{therpum}), (\ref{ptot}), (\ref{pumpedheatlambda}) and (\ref{worklambda}) which show that the pumped heat, the concomitant heat-work conversion and the dissipated power have a geometric interpretation. In the same section we will also analyze several classes of adiabatic machines depending on the various adiabatic drivings. Following this general formulation, 
Section.~\ref{Examples} focuses on two specific examples of thermal machines, which are particularly relevant for experimental implementations. We first consider a driven qubit which is asymmetrically and weakly coupled to two bosonic thermal baths. We then discuss a driven quantum dot coupled to two electron reservoirs. Conclusions and some additional perspectives related to our work are presented in Section~\ref{conclusions}. The appendices contain further details 
on the derivation of the main results of the paper and explicit calculations for the examples presented in the main text.

\section{Model of a geometric thermal machine}
\label{model-heat-engine}

A sketch of the geometric thermal machine that we analyze throughout this paper is shown in Fig.~\ref{generic-q-machine}. It consists of a central region containing the working substance, constituted by a few-level quantum system, coupled to two thermal baths. The quantum system is periodically driven by a set of $N$ slowly-varying parameters $\vec{X}(t)$. 
The baths are macroscopic reservoirs of bosonic excitations or fermionic particles. The macroscopic variables characterizing the thermal environment such as the bath temperatures can also slowly vary in time. We parametrize the bath temperatures as  $T_\alpha(t) = T + \delta T_{\alpha}(t)$ (with $\alpha=L,R$ referring to the left and right reservoirs) and define $\Delta T(t)= \delta T_L(t) - \delta T_R(t)$. A (possible) time dependence in the bath temperatures is only included in $\delta T_{\alpha}(t)$. We assume that the right reservoir $R$ is the colder one. 

Let us start with the simple observation that the thermal bias, without the action of the ac driving, induces a net heat flow  from the hot to the cold reservoir. On the other hand, it is useful to consider an analogy with the operation of classical machines and notice that 
 the modulation of the parameters $\vec{X}(t)$
is introduced by some mechanism, which is akin to a weight moving a wheel in the classical case.
By the combined effects of thermal bias and ac driving forces,
it is possible to realize {\em heat-work conversion}, which constitutes the 
 key for the operation of the device as a thermal machine. Two main operational modes are possible. (i) In the heat engine mode, part of the heat flowing in the direction of the thermal bias is transformed into work performed against the mechanisms 
 ruling the dynamics of $\vec{X}(t)$. (ii) In the refrigerator mode, part of the work induced by the action of the ac parameters can be used to extract heat from the cold reservoir, against the action of the thermal bias. In the latter case, the
 thermal bias plays the role of the weight. In the operation of the thermal machines, these processes come along with dissipation of energy leading to entropy production. The efficiency of the thermal machine relies on the appropriate balance between the heat-work conversion mechanism and dissipation.

\begin{table}[t]
\begin{tabular}{|l|l|}
\hline
N \;\;\; 			&    	Number of slowly varying coupling parameters \\ \hline
N + 1 \;\;\; 			&    	Number of slowly varying coupling parameters \\  
 	 \;\;\; 			&    	including thermal bias \\ \hline
$\vec{v} \;\;\; $ 		&  	Arrows denote $N$-dimensional vectors  \\ \hline
$\bf{v}$ \;\;\;  		&     	Bold fonts denote $(N+1)$-dimensional vectors  \\ \hline
$\ell \;, \ell'$ \;\;\;  	&    	Labels of elements of $N$-dimensional vectors \\ 
\;\;\; 			&       or matrices \\ \hline
$\mu \;, \nu$	\;\;\;	&     	Labels of elements of $(N+1)$-dimensional  \\ 
     \;\;\; 	   		& vectors or matrices \\ \hline
$ \overleftrightarrow{M} $ 	\;\;\;	&     	$N\times N$ matrix \\ \hline
$\underline{ \underline{{\bf M}}}$ & $(N +1)\times (N+1)$ matrix \\ \hline

\end{tabular}
\label{notation-table}
\caption{Notation used in the text}
\end{table}

\subsection{Heat, work, and operational modes}
As we are interested in the dynamics for slow driving and small temperature biases, it is convenient to define the $N+1$-dimensional vector of "velocities",
\begin{equation}
	\dot{{\bold X}}(t) = \left\{ \dot{\vec{X}}(t), \Delta T(t)/T \right\}
\label{x-vec}
\end{equation}
These two types of vector notation (arrow and bold character) appear in several places throughout the paper. For later reference, the Table~ I summarizes the different symbols used in the text. 

A temperature bias as well as time-dependent system and bath parameters generally induce net heat transport between the reservoirs. At the same time, any driving mechanism generates heat that is dissipated into the reservoirs. Hence, the total heat current entering a given reservoir has a component resulting from the net transport between the two reservoirs and a component originated in the dissipation  because of the action of the driving forces. The net heat current $J^Q_{\alpha}$, averaged over one cycle of period $2\pi/\Omega$,  satisfies  \cite{lilimos},
\begin{equation}\label{dis}
J^Q_{\rm L}+J^Q_{\rm R}=P,
\end{equation}
where  $P$ is the total dissipative power generated by the driving forces, also averaged over one period. Identifying the component due to transport and that due to dissipation in $J^Q_{\alpha}$ is  a non trivial task in general. The transport component satisfies
\begin{equation}\label{net}
 	J^{Q}_{\rm tr, R}=- J^{Q}_{ \rm tr , L} \equiv J^{Q}_{ \rm tr},
\end{equation}
and we notice that only the total dissipative heat contributes to Eq. (\ref{dis}).
In the next section, we exactly calculate  $J^Q_{\alpha}$  to linear order in $\dot{{\bold X}}(t)$ and we show that it satisfies Eq. (\ref{net}). Hence, we identify it with the leading term of the transport current. 

The net heat transported per cycle between the two reservoirs is
\begin{equation}\label{qtra}
Q_{\rm tr} = \frac{\Omega}{2\pi} J^{Q}_{ \rm tr}.
\end{equation}
This component is defined such that  $Q_{\rm tr} >0$ when heat flows in the direction of the thermal bias (hot to cold). 
We also define the net work $W$ performed on the system by the ac forces during one cycle. 
We take $W>0$ when the ac forces exert work on the system. 
The balance between $Q_{\rm tr}$ and $W$  is the key to the performance of the thermal machine, which may operate as a heat engine by transforming heat into work against the time-dependent driving or as a refrigerator, by using the work performed by the $ac$ driving to pump heat from the cold to the hot reservoir.
In the absence of heat-work conversion, one finds that both $Q_{\rm tr}\geq 0$ and $W \geq 0$.
In the heat-engine mode, the heat--work conversion mechanism operates against the ac forces and consequently $W<0$. In the refrigerator
mode, the heat--work conversion mechanism operates by using part of the work done by the ac forces to pump heat against the thermal bias, so that $Q_{\rm tr}<0$.

It is straightforward to generalize our considerations to multi-terminal devices or to include additional macroscopic variables beyond temperature such as an electrochemical potential difference between reservoirs.

\subsection{Adiabatic linear response}
\label{linear_ad}

To analyze the performance of the adiabatic thermal machines, we need to compute the currents. This can be done by conventional many-body techniques, such as the non-equilibrium Green's function formalism, scattering matrix theory (for systems without many-body interactions), or master equations (for weak coupling between system and reservoirs). Although we use these techniques to solve specific examples, we employ a Hamiltonian representation for the temperature difference and a Kubo linear response framework for small $\Delta T$
to derive general results. This enables us to analyze the energy dynamics induced by the thermal driving on the same footing with that induced by the time-dependent driving. Here we follow  Luttinger's approach~\cite{luttinger} to thermal transport which 
introduces a "gravitational" potential whose gradients induce energy flows akin to the electrical currents induced by gradients of the electrochemical potential. Details of this approach are given in Appendix~\ref{Lutt-theory}. 

We then introduce the Hamiltonian ${\cal H}$ governing the system of Fig.~\ref{generic-q-machine}, which can be expressed as 
\begin{equation}
  	{\cal H}(t)= {\cal H}_S(t) + {\cal H}_{\rm baths} + {\cal H}_{c}  +  {\cal H}_{\rm th}(t).
  	\label{hamtot}
\end{equation}
The first term ${\cal H}_S(t)$ is the Hamiltonian of the quantum system. It depends on time through the $N$ slowly and periodically varying parameters (driving potentials) $\vec{X}(t)= \left\{X_{\ell}(t)\right\}$ {with} $\ell = 1,\dots,N$, so that ${\cal H}_S(t) \equiv {\cal H}_S[\vec{X}(t)]$. The second term describes the two reservoirs $ {\cal H}_{\rm baths} = {\cal H}_{\rm R}+{\cal H}_{\rm L}$, which are macroscopic systems of bosonic excitations or fermionic particles. In the latter case, they are held at the same chemical potential $\mu_L=\mu_R=\mu$ and should be described by the grand-canonical Hamiltonian, ${\cal H}_{\alpha} \rightarrow {\cal H}_{\alpha} -\mu {\cal N}_{\alpha}$, where ${\cal N}_{\alpha}$ denotes the number of particles in reservoir $\alpha$. The coupling between system and reservoirs, such as tunneling of particles and/or the exchange of energy between system and reservoirs, is captured by ${\cal H}_{c}$. Its form depends on the specific implementation, and some examples will be described later in the paper. The last term in Eq.~(\ref{hamtot}) accounts for the fact that the two reservoirs are held at  different temperatures and derives from the Luttinger formulation of thermal transport. Adapting the definition of Eq.\ (\ref{at}) to the present case, we define
\begin{equation}
	\label{hth}
	{\cal H}_{\rm th}(t) = - \sum_{\alpha=L,R} {\cal J}^E_{ \alpha}(t) \xi_{\alpha}(t),
\end{equation}
where $\xi_{\alpha}(t)$ plays the same role as the thermal vector potential and the operator representing the energy flux entering reservoir $\alpha$ is given by
\begin{equation}
	{\cal J}^E_{ \alpha}=  \dot{\cal H}_{\alpha}=-i \left[{\cal H}_{\alpha}, {\cal H} \right]/\hbar. 
\end{equation} 
Here, ${\cal H}_{\alpha}$ is the Hamiltonian of reservoir $\alpha$. When the chemical potential is the same for all reservoirs, time averaging the mean value of this operator over one period $\tau=2\pi/\Omega$ directly gives the heat current,
\begin{equation}
	J^Q_{\alpha}= \frac{\Omega}{2\pi} \int_0^{2\pi/\Omega} \mathrm{d}t \langle {\cal J}^E_{ \alpha}(t) \rangle.
\end{equation}
The relation between the Luttinger field and the temperature bias, the counterpart of Eq.\ (\ref{adel}), reads
\begin{equation}\label{dotpsi}
\dot{\xi}_{\alpha}(t)=  \delta T_{\alpha}(t)/T .
\end{equation}

Our quantum machine operates in a regime in which both the driving parameters $\vec{X}(t)$ and the temperature bias $\delta T_{\alpha}$ (with the associated parameter $\xi_{\alpha} (t)$) vary in time.  Adiabatic driving implies that the driving frequency $\Omega$ is small compared to any characteristic frequency of the system's degrees of freedom as well as the relevant relaxation times associated with the coupling to the reservoirs. We can then regard the velocities at which the parameters are changed  and the temperature bias
as sufficiently small so that the currents can be computed in linear response in $\dot{\bf X}$. This  procedure was previously introduced in Ref. \onlinecite{ludovico} and it is similar to the one of Ref. \onlinecite{anatoli1} for closed driven systems. The adiabatic time evolution of any observable ${\cal O}$ is described by the Kubo-like formula
\begin{eqnarray}
	\label{linearo}
 	\langle {\cal O}\rangle(t) = 
  	\langle  {\cal O} \rangle_t &+& \sum_{\ell=1}^{N} \chi^{\rm ad}_t\left[{\cal O},{\cal F}_{\ell}\right] \dot{X}_{\ell} (t) \nonumber \\
	&+& \sum_{\alpha=L,R} \chi^{\rm ad}_t\left[{\cal O},{\cal J}^E_{ \alpha}\right] \dot{\xi}_{\alpha} (t).
\end{eqnarray}
Here, the left-hand side denotes an average with respect to the nonequilibrium density matrix, while $\langle  {\cal O} \rangle_t$ is an average with respect to the equilibrium density matrix of the frozen Hamiltonian ${\cal H}_t={\cal H}_S(t) + {\cal H}_{\rm baths} + {\cal H}_{c} $,  $\rho_t=\sum_m p_m |m \rangle \langle  m|$, where $p_m= e^{-\beta \varepsilon_m}/Z_t$, $\beta=1/k_B T$, and ${\cal H}_t |m\rangle =\varepsilon_m |m\rangle$.  Notice that the instantaneous eigenvectors $|m\rangle$ and eigenenergies $\varepsilon_m$ depend on the time $t$. Here we have introduced the operator 
\begin{equation}
	{\cal F}_{\ell} = - \frac{\partial {\cal H}}{\partial X_{\ell}}, \;\;\;\; {\rm with} \;\; \ell=1,\dots,N
\end{equation}
which has the interpretation of a force induced by the driving. The {\em adiabatic response functions} appearing in Eq.~(\ref{linearo}) take the form  
\begin{equation}\label{chiad}
   	\chi^{\rm ad}_t\left[{\cal O}_1,{\cal O}_2  \right] = - \frac{i}{\hbar} \int_{-\infty}^t \mathrm{d}t^{\prime} (t-t^{\prime}) \langle \left[ {\cal O}_1(t),{\cal O}_2(t^{\prime})\right]\rangle_t.
\end{equation}
We have also assumed that the perturbations are switched on at $t_0=-\infty$. 

Within this framework, we can evaluate the adiabatic evolution of any observable. We are particularly interested in the energy current flowing into the coldest reservoir and the induced forces. Similar to the definition in Eq.~(\ref{x-vec}), we find it convenient  to define the  $N+1$-dimensional force vector
 \begin{equation}\label{cal f}
	 \mathbfcal{F} = (\vec{{\cal F}} , {\cal J}^E_R ).
\end{equation}
Using this notation, the adiabatic dynamics for the forces and the energy current into the coldest reservoir can be written as
\begin{equation}
  	\langle \mathbfcal{F}  \rangle(t) =  \langle  \mathbfcal{F}  \rangle_t +  \underline{ \underline{{\bf \Lambda}}}({\vec X}) \cdot \dot{{\bf X}}.
\label{linearof}
\end{equation} 
As expected, the physical response depends on the two Luttinger parameters $\xi_L(t)$ and $\xi_R(t)$ only through the temperature bias $\dot{X}_{N+1}(t)= \Delta T(t)/T$, as can be seen using Eqs.~(\ref{ident}) and (\ref{fj}). In Eq.~(\ref{linearof}), we introduce the response matrix $\underline{ \underline{{\bf \Lambda}}}({\bf X})$ with elements defined as 
\begin{equation}
\label{lambda}
\Lambda_{\mu,\nu}(\vec{ X}) = 
	\Bigg\{\begin{matrix} 
		\chi_t^{\rm ad} \left[ {\cal F}_\mu , {\cal F}_{\nu} \right] & \mu \le N 
		\\ \\
		\sum_{\alpha =L,R} \chi_t^{\rm ad} \left[ {\cal J}^E_{\alpha},{\cal F}_\nu  \right] & \mu = N+1\\
		\end{matrix}
\end{equation} 
Note  that in deriving the linear response expression for the current, one should neglect the term ${\cal H}^{\rm th}_t$, which would lead to a "diamagnetic" component of the heat current \cite{tatara}. The notation in Eq.~(\ref{lambda}) highlights the fact that the $\Lambda_{\mu,\nu}(\vec{ X})$ depend on time only through the parameters $\vec{X}$.  

As the coefficients of Eq.~(\ref{lambda}) are evaluated with respect to the frozen equilibrium density matrix, they obey the Onsager relations \cite{ludovico,cohen}
\begin{eqnarray}
	\label{onsager}
	\Lambda_{\mu,\nu}(\vec{ X},\vec{B})&=& s_{\mu} s_{\nu} \Lambda_{\nu,\mu}(\vec{ X},- \vec{B}) ,
\end{eqnarray}
where $s_{\nu}=\pm$ for operators ${\cal F}_{\nu}$ which are even/odd under time reversal. In view of its relevance for time-reversal symmetry, we made a possible dependence on an applied magnetic field $\vec{B}$ explicit here, but will suppress it in the following unless necessary.

\subsection{Adiabatic forces, currents, and entropy production over a cycle}
\label{ener-bal}

In the geometric description of the adiabatic thermal machines, the central role is played by integrals of the forces in Eq.\ (\ref{linearof}) over a period, rather than by the instantaneous quantities. First consider the energy current 
$\langle {\cal J}_R^E \rangle(t)$ which leads  to a description of the heat fluxes introduced in Eq. (\ref{net}) within the adiabatic linear response formalism. The average of the instantaneous heat current over one period, $\langle {\cal J}_R^E \rangle(t)$, defines the transported heat flux within the adiabatic linear response formalism. In fact, evaluating this current  with the adiabatic expansion of Eq.~(\ref{linearo}) and using 
 the identities of Eqs.~(\ref{ident}) and (\ref{fj})  we can see that the average over one period is identical in magnitude and opposite in signs at the two reservoirs. Hence, we eliminate the label $\alpha$ and write
  \begin{eqnarray}\label{therpum}
	J^{Q}_{\rm tr} &=&
 \frac{\Omega}{ 2\pi }\int_0^{2\pi/\Omega} \mathrm{d}t \; \sum_{\nu=1}^{N+1}  \Lambda_{N+1,\nu}(\vec{ X})  \dot{X}_{\nu} (t).
 \end{eqnarray}
 The term corresponding to the sum $\nu=1, \ldots, N$ 
 is the pumping contribution to the heat current. The literature on pumping of charge and heat, starting with the seminal paper by Thouless~\cite{Thouless1983}, is so vast that it would be impossible to give a proper account of it. A brief overview can be found in the reviews~\cite{Avron2003,Xiao2010}. One of the key results of the present paper is to show how pumping affects the operation of a quantum thermal machine, thus paving the way to observe geometric effects in the operating mode of these systems. The last term of Eq. (\ref{therpum}), corresponding to $\nu=N+1$, is the heat current flowing in response to a finite temperature bias across the device. 

For a single driving parameter and $\Delta T=0$, it is straightforward to show that the pumped heat current  vanishes. At least two parameters are necessary for pumping. This was originally noticed in the framework of scattering matrix theory for driven electron systems \cite{avron,brouwer}. Moreover, a spatially symmetric system has $\chi_t^{\rm ad} \left[ {\cal J}^E_L, {\cal F}_{\ell} \right] = - \chi_t^{\rm ad} \left[ {\cal J}^E_R, {\cal F}_{\ell} \right]$, so that these quantities should be zero in view of Eq.~(\ref{fj}). Hence, breaking of spatial symmetry is another necessary condition for a non-vanishing pumping contribution to the heat current \cite{mobu,lilimos}. 

The net generated power has components associated to the time-dependent driving forces as well as to the thermal bias,
\begin{widetext}
\begin{eqnarray}
	\label{ptot}
	P &=&  \frac{\Omega}{ 2\pi } \int_0^{2\pi/\Omega}  \mathrm{d}t \left(\sum_{\ell=1}^N \langle {\cal F}_\ell \rangle \dot{X}_\ell (t) + \sum_{\alpha,\beta=L,R}   \langle {\cal J}^E_{\alpha} \rangle(t)  \dot{\xi}_{\alpha}(t)  \right)
	 =   \frac{\Omega}{ 2\pi }  \int_0^{2\pi/\Omega}  \mathrm{d}t  \;\; \dot{{\bf X}}  \cdot \underline{ \underline{{\bf \Lambda}}}({\vec X}) \cdot \dot{{\bf X}},
\end{eqnarray}
\end{widetext}
The response matrix on the right-hand side of Eq.~(\ref{ptot}) was introduced through the definition of forces and the energy current in Eq.~(\ref{linearof}). While Eqs.~(\ref{therpum}) for the fluxes are linear in $\dot{{\bf X}}$, Eq.~(\ref{ptot}) is bilinear in these parameters. This reflects the fact that the dissipated heat, defined in Eq.~(\ref{net}) is at least second order in these quantities -- equivalent to 
being ${\cal O}(\Omega^2)$ \cite{mobu,lilimos}. The cross terms proportional to the thermal bias and {\em ac} driving usually have opposite signs and cancel one another when evaluating the total power. This happens, in particular, in the absence of a magnetic field with driving forces symmetric under time reversal, as a consequence of the Onsager relations (\ref{onsager}).

From Eq.~(\ref{dis}) for the total dissipated heat flux we have the following expression for the entropy production rate
\begin{equation}\label{s}
T \dot{S}= J^{Q}_{L}+J^Q_R=P.
\end{equation}
Substituting Eq.~(\ref{ptot})  we get
\begin{equation}\label{sdot0}
\dot {S}= \frac{\Omega }{2 \pi T} \int_0^{2 \pi/\Omega} \mathrm{d}t  \;\; \dot{{\bf X}} (t) \cdot \underline{ \underline{{\bf \Lambda}}}({\vec X}) \cdot \dot{{\bf X}}(t).
\end{equation}
We  present an alternative derivation for the above expression in Appendix \ref{entropy-rate}.

The forces $\langle {\cal F}_\ell \rangle(t)$ enter the work performed by the thermal machine, as will be discussed in more detail in Sec.\ \ref{sec:thermalgeo} below. We also find it useful to introduce average of the force over one period, 
\begin{equation}
F_{\ell}=\frac{\Omega}{2\pi}  \int_0^{2\pi/\Omega}  \mathrm{d}t \langle {\cal F}_{\ell} \rangle(t)= F_{\ell,\rm BO}+F_{\ell,\rm ar}, \;\;\;\ell,=1,\ldots, N.
\label{timeaverageforce}
\end{equation}
The first term of Eq. (\ref{timeaverageforce}) corresponds to the instantaneous {\em equilibrium} (Born-Oppenheimer) description given by the first term of Eq. (\ref{linearof}), while the second term is the first order {\em adiabatic reaction force} defined in Ref. \onlinecite{berry2010}. 

\section{Geometric characterization}
\label{geoF}

\subsection{Thermal geometric tensor}
\label{geo-section}

It is instructive to decompose the tensor $\Lambda_{\mu,\nu}(\vec{ X})$ into its symmetric and antisymmetric parts, 
\begin{equation}
  \Lambda^{S,A}_{\mu,\nu} = \frac{1}{2}\left( \Lambda_{\mu,\nu} \pm \Lambda_{\nu,\mu}\right).
\end{equation}
Equation (\ref{sdot0}) for the entropy production implies that the symmetric component $\Lambda^S_{\mu,\nu}$ controls dissipation. Since the rate of entropy production $\dot{S}$ is non-negative, the symmetric part  $\Lambda^S_{\mu,\nu}$ can be viewed as a metric tensor on the space of thermodynamic states \cite{crooks,reza,marti}. Then, geodesics with respect to this metric correspond to 
adiabatic trajectories which minimize dissipation \cite{crooks,reza,marti}. This contribution to $\Lambda_{\mu,\nu}(\vec{ X})$ has also been referred to as geometric friction \cite{Berry1993,crooks,campisi2}. 

We can obtain an explicit expression for $\Lambda_{\mu,\nu}$ from the Lehmann representation (see details in App.~\ref{lehmann}). The result for the symmetric component is
\begin{eqnarray}
\label{lambfs}
\Lambda^S_{\mu,\nu}(\vec{ X}) = & & \hbar \pi \lim_{\omega \rightarrow 0} \sum_{n,m} p_m  
\frac{(\varepsilon_n-\varepsilon_m)^2}{\omega}\mbox{Re}[\langle \partial_{\mu} m|n\rangle \langle n| \partial_{\nu} m\rangle ]\nonumber \\
& & \times \left[\delta(\omega-(\varepsilon_m-\varepsilon_n))- \delta(\omega-(\varepsilon_n-\varepsilon_m))\right].
\end{eqnarray}
Here, $|m\rangle$ and $\epsilon_m$ denote the instantaneous eigenstates and eigenenergies of ${\cal H}_t$ and $p_m$ is the corresponding thermal weight, with the same definitions  as in Eq.\ (\ref{linearo}). Similarly, the antisymmetric component can be expressed as
\begin{eqnarray}\label{lambfa}
\Lambda^A_{\mu,\nu}(\vec{ X}) = 2 \hbar\sum_m p_m \;\mbox{Im} \left[\langle \partial_{\mu} m | \partial_{\nu} m \rangle\right].
\end{eqnarray}
In the limit of zero temperature, the sum over $m$ is dominated by the ground state and  $\Lambda^A_{\mu,\nu}(\vec{ X})$ reduces to its Berry curvature. For $\Delta T=0$, this component can be viewed as a velocity-dependent force, akin to a Lorentz force, which does not contribute to the net entropy production. This contribution has been referred to as geometric magnetism \cite{Berry1993,hanggi,Bode2011,Bode2012,Thomas2012}. 

It is interesting to compare $\Lambda_{\mu,\nu}$ to the quantum geometric tensor 
for the instantaneous ground state $|\psi\rangle$ of a closed system as a function of parameters $X_\ell$ \cite{anatoli2,anatoli3},
\begin{equation}
     g_{\ell,\ell^{\prime}} = \left\langle {\partial_\ell \psi} \left |  {\partial_{\ell^{\prime}} \psi}\right. \right\rangle - \left\langle \left. {\partial_\ell \psi}\right |  \psi \right\rangle\left\langle  \psi\left |  {\partial_{\ell^{\prime}} \psi}\right. \right\rangle.
\end{equation}
 Analogous to $\Lambda_{\mu,\nu}$, the symmetric part of $g_{\ell,\ell^{\prime}}$ defines a metric on the manifold of ground states and the antisymmetric part equals the Berry curvature. The crucial difference between the two tensors is that the quantum geometric tensor is defined for a discrete spectrum, while $\Lambda_{\mu,\nu}$ assumes a continuous spectrum. This does not lead to essential differences for the antisymmetric components of the tensors which are non-dissipative. In contrast, the symmetric part of $\Lambda_{\mu,\nu}$ controls dissipation and therefore vanishes for a discrete (or gapped) spectrum. We can therefore view $\Lambda_{\mu,\nu}$ as the analog of the quantum geometric tensor for systems with continuous spectra. In view of this analogy, we refer to $\Lambda_{\mu,\nu}$ as the {\em thermal geometric tensor}. 

In  time reversal symmetric systems subject to driving parameters $\vec X$ which also respect time reversal symmetry, different parts of the thermal geometric tensor are either purely symmetric or antisymmetric. The Onsager relations (\ref{onsager}) imply that $\Lambda_{\ell,\ell^{\prime}} = \Lambda_{\ell^{\prime},\ell}$  ($\ell,\ell^{\prime} = 1, \ldots, N$) is purely symmetric (corresponding to geometric friction without geometric magnetism). In contrast, $\Lambda_{N+1,\ell} = - \Lambda_{\ell, N+1}$ (corresponding to geometric magnetism without geometric friction). In systems which break time reversal symmetry, both the symmetric and the antisymmetric components of the thermal geometric tensor are  generally nonzero. 
 
\subsection{Thermal machines and geometry}
\label{sec:thermalgeo}

The above analysis implies that there are several purely geometric quantities which enter into the operation of adiabatic quantum thermal machines. We will show in the following that in a very concrete sense, it is the geometric aspects (in the sense of Berry) which are responsible for the heat-work conversion underlying thermal machines. The Carnot limit of the efficiency is reached in a purely geometric thermal machine, and deviations from the Carnot limit are due to nongeometric contributions.

An essential quantity is the total heat transported between the leads per cycle, $Q_{\rm tr} $ defined in Eq. (\ref{qtra}). In a heat engine, this heat is in part converted into useful work while in a refrigerator, this heat is extracted from the colder reservoir. The transported heat takes the form
\begin{equation}
  Q_{\rm tr} = \oint \sum_{\ell=1}^N \Lambda_{N+1,\ell} {\mathrm d}X_\ell + \oint {\mathrm d}t    
        \Lambda_{N+1,N+1} \frac{\Delta T}{T}.
\label{pumpedheatlambda}
\end{equation}
The first term on the right hand side is geometric, depending only on the path, and has a simple physical interpretation. It is just the heat 
which is  pumped between the reservoirs due to the periodic variation of the parameters $\vec X$,   
\begin{equation}
   Q_{\rm tr,ac} = \oint \sum_{\ell=1}^N \Lambda_{N+1,\ell} {\mathrm d}X_\ell .
\label{pumpedheatlambda2}
\end{equation}
The second term describes the heat current driven by the applied temperature bias as a result of the heat conductance $\Lambda_{N+1,N+1}$ of the system. Notice that the two terms typically have a different dependence on the period $2\pi/\Omega$. Due to its geometric nature, the first term is independent of the period. In contrast, the second term is in general proportional to the period. 

The pumped heat per cycle is essential for the operation of adiabatic quantum thermal machines. To see this, we compute the work $W = \oint {\mathrm d}{\vec X}\cdot{\vec F}$ per period performed on the system during one cycle of the {\em ac} sources. The forces, as described by Eq. (\ref{linearof}), have an instantaneous and a linear-response component. The instantaneous contribution depends only on the parameters $\vec X$ and is evaluated in the absence of the temperature bias. This {\em equilibrium} contribution to the force is necessarily conservative (in the mechanical sense) and thus gives a vanishing contribution to the work performed over a cycle. Thus, only the linear-response component contributes to the work per cycle, 
\begin{equation}
   W = \oint {\mathrm d}t \sum_{\ell,\ell'=1}^N {\dot{X}_\ell} \Lambda_{\ell,\ell'}{\dot{X}}_{\ell'} + \oint \sum_{\ell=1}^N {\mathrm d}X_\ell \Lambda_{\ell,N+1} \frac{\Delta T}{T}.
\label{worklambda}
\end{equation} 
First consider the second term on the right hand side. For constant $\Delta T/T$, this term is again a purely geometric line integral over a closed contour. Unlike the contribution of the instantaneous component, this term is in general nonconservative and gives a finite contribution when integrated over a closed cycle. The reason is that this term originates from the {\em nonequilibrium} contribution to the force which is generated by the temperature bias. Along with Eq.\ (\ref{pumpedheatlambda2}) for the pumped heat, this geometric term is the essence of {\em heat-work} conversion and hence crucial for the operation of the thermal machine. In contrast, the first term in Eq.\ (\ref{worklambda}) describes frictional losses. Unlike the second term, which can take either sign, this term is always positive. It then becomes evident that heat-work conversion is rooted in the geometric terms in Eqs.\ (\ref{pumpedheatlambda}) and (\ref{worklambda}), and it is the nongeometric terms (in the sense of Berry) that are responsible for losses. We will see this again below when we discuss the efficiencies of quantum thermal machines.

As a result of the Onsager relations (\ref{onsager}), the geometric contributions to the transported heat and the work 
are very closely related. If the system is time reversal invariant (which also requires that the parameters ${\vec X}$ couple to time-reversal-even operators), the Onsager relations imply that $\Lambda_{N+1,\ell} = - \Lambda_{\ell,N+1}$ and the prefactor of $\Delta T/T$ in Eq.\ (\ref{worklambda}) just equals minus the pumped heat between the reservoirs. We can then understand the operation of a heat engine as follows. During one cycle of the machine, the cyclic variation of the parameters pumps heat from the high-temperature to the low-temperature reservoir. The corresponding change in free energy is converted into work $W$ performed on a load (i.e., $W<0$). Here, the load corresponds to an external agent which couples to the dynamics of the parameters $\vec{X}$. This is analogous to the operation principle of inverted quantum pumps as adiabatic quantum motors \cite{Bustos2013,torque4,motor1,motor2,motor3}. Similarly, in a refrigerator work $W=- Q_{\rm tr,ac}\Delta T/T >0 $ must be supplied by the {\em ac} sources to overcome the thermal bias and to pump heat $Q_{\rm tr,ac}$ from the low-temperature to the high-temperature reservoir.  

It is also interesting to discuss this heat-work conversion in the context of the entropy production rate defined in Eq. (\ref{s}). With the definitions of this section, we can write
\begin{equation}\label{sqw}
T\dot{S}= \frac{\Omega}{2 \pi} \left(W + Q_{\rm tr} \frac{\Delta T}{T}  \right).
\end{equation}
The first term corresponds to the total power generated by the ac sources, while the second term corresponds to the power invested to transport the heat $Q_{\rm tr}$ per cycle in the presence of the thermal bias $\Delta T$. Due to the heat-work conversion, the geometric component of $W$ exactly cancels the component $Q_{\rm tr,ac}$ of $Q_{\rm tr}$ in the dissipated power (still assuming time-reversal invariance). Entropy production is then associated with the non\-geo\-metric contributions to heat and work. In a heat engine, a negative balance of the two terms contributing to Eq.\ (\ref{worklambda}), $W<0$, can be used to work against the load. In a refrigerator, both terms are positive since one has to overcome the frictional losses in addition to pumping heat from the cold to the hot reservoir. It is important to notice that the two terms in Eq.\ (\ref{worklambda}) are typically of different orders in the period $2\pi/\Omega$. While the first, nongeometric contribution is inversely proportional to the period, the second, geometric contribution is independent of it. Thus, one can often neglect the nongeometric term when considering the limit of small frequency $\Omega$. As we will show below, we note that under certain circumstances the first term in Eq.\ (\ref{worklambda}) can also be viewed as a geometric quantity even though it cannot be immediately rewritten as a line integral.

We close this section by a few additional remarks. The operation of a heat engine or refrigerator requires that a net amount of heat $Q_{\rm tr,ac}$ is pumped between the reservoirs during a cycle, requiring that the force is nonconservative. Above, we have focused on the case that $\Delta T/T$ is constant over the cycle. In principle, the conditions for the operation of adiabatic quantum thermal machines can be less stringent if one allows $\Delta T/T$ to vary along the cycle, for instance by coupling the system to different reservoirs at different stages. 

In the absence of time reversal symmetry, the Onsager relations connect the response functions $\Lambda_{\mu,\nu}$ at different magnetic fields. In this case, there is no general relation between $\Lambda_{N+1,\ell}$ and $\Lambda_{\ell,N+1}$ for a fixed magnetic field, and in addition to the antisymmetric contribution $\Lambda^A_{N+1,\ell} = - \Lambda^A_{\ell,N+1}$, there could also be a symmetric contribution, $\Lambda^S_{N+1,\ell} =  \Lambda^S_{\ell,N+1}$. Unlike $\Lambda_{\mu,\nu}^A$, the symmetric $\Lambda_{\mu,\nu}^S$ is associated with entropy production and dissipation according to Eq.\ (\ref{ptot}). Even if both the dissipative and the nondissipative contributions to the pumped heat flow from the hot to the cold reservoir, the work performed on a load would involve the difference between the antisymmetric and the symmetric contribution. 

The time average of the forces $\vec F$ as defined in Eq.\ (\ref{timeaverageforce}) also has contributions which are purely geometric.   From Eq.~ (\ref{linearof}), the first-order adiabatic reaction component can be readily rewritten as
\begin{equation}
F_{\ell, {\rm ar}} = \frac{\Omega}{2\pi}\left\{ \oint \sum_{\ell'=1}^N\Lambda_{\ell,\ell'} {\mathrm d}{X}_{\ell'}  + \oint {\mathrm d}t \Lambda_{\ell,N+1} \frac{\Delta T}{T} \right\}
\label{avforceline}
\end{equation}
Here, the first term on the right hand side is a line integral which is purely geometric in that it depends only on the path. 

Finally, we remark that under certain conditions, the dissipated component of  $W$, corresponding to the first term of Eq. (\ref{worklambda}), can also be formally represented in terms of a line integral over a closed path in parameter space. This is not as straightforward as for Eqs.\ (\ref{avforceline}), (\ref{pumpedheatlambda}), and (\ref{worklambda})  since the power is bilinear in $\dot{\bf X}$. It is, however,  possible when there exists a well-defined mapping between $\dot{{\bf  X}}$ and ${\bf X}$ as the latter varies along the closed path $\gamma$. In particular, such a mapping exists for the case of periodic driving. For a smooth path $\gamma$, one can write the relations $\dot{X}_{\mu} = \Omega g_{\mu}( \vec{ X})|_{\gamma}$ for all $\mu$, where the functions $g_{\mu}(\vec{ X} )|_{\gamma}$ are defined by eliminating the parametrization in $t$ between $X_{\mu}(t)$ and $\dot{X}_{\mu}(t)$. Then, we can write the dissipated power as a line integral by using this relation to eliminate one of the factors of $\dot{X}_{\mu}$ in Eq.\ (\ref{ptot}) via these relations. Note that the resulting line integral has a prefactor of $\Omega$, making it explicit that the dissipated power is inversely proportional to the period of the driving, as already mentioned above. 

The line integrals controlling the operation of adiabatic thermal quantum machines are reminiscent of line integrals over Berry connections. This motivates us to introduce the vector fields
\begin{equation}\label{amu}
\vec{A}^{A/S}_{\mu}=\left( \Lambda^{A/S}_{\mu, 1}(\vec{ X}), \ldots, \Lambda^{A/S}_{\mu, N}(\vec{ X})\right)
\end{equation}
with $\mu=1, \ldots, N+1$ for the rows of the thermal geometric tensor. 
Similarly, we introduce 
\begin{eqnarray}\label{atildemu}
 \tilde{\vec{A}}&=& \sum_{\ell} \left(\tilde{ \Lambda}^S_{\ell,1}(\vec{X}), \ldots, \tilde{\Lambda}^S_{\ell,N}(\vec{X})\right),
\end{eqnarray}
where $\tilde{\Lambda}^S_{\mu, \nu}(\vec{ X} )=g_\mu(X_{\mu}) \Lambda^S_{\mu, \nu}(\vec{X})$.
 These vector fields control the pumped heat and the work performed on the system as well as the dissipated power. Thus, they are useful to illustrate the operation of the specific thermal machines which we discuss in Sec.\ \ref{Examples}.  In terms of these vector potentials Eqs. (\ref{pumpedheatlambda2}) and (\ref{worklambda}) read, respectively,
 \begin{equation}\label{q-a}
 Q_{\rm tr, ac}=\oint  \vec{A}_{N+1}(\vec{X})  \cdot d \vec{X},
 \end{equation}
 with $\vec{A}_{\mu}(\vec{X}) = \vec{A}^A_{\mu}(\vec{X}) +\vec{A}^S_{\mu}(\vec{X}) $
and
 \begin{equation}\label{w-a}
 W=\oint  \left[ \tilde{\vec{A}}(\vec{X}) - \frac{\Delta T}{T} \left(\vec{A}^A_{N+1}(\vec{X}) - \vec{A}^S_{N+1}(\vec{X}) \right)\right] \cdot d \vec{X}.
 \end{equation}
 In the latter equation, the last term does not contribute for many systems. In particular, this is the case in the presence of time-reversal symmetry (including driving parameters $\vec{X}$ coupling to time-reversal-even operators). In such cases,  we can write $ W=\oint   \tilde{\vec{A}}(\vec{X})  - (\Delta T/T) Q_{\rm tr, ac}$.
 
\subsection{Efficiencies}

\subsubsection{Heat engine}
\label{eff_heat}
In a heat engine, heat transported from the high to the low temperature reservoir is partially converted into useful work. We can then define an efficiency  for the heat engine as 
\begin{equation}
\eta^{\rm (he)} = \frac{-W}{Q_{\rm tr}}.
\label{effx_he}
\end{equation}
This expression can be readily analyzed for a time-reversal-invariant system with constant $\Delta T/T$. In the limit of adiabatic operation of the heat engine, $\Omega\to 0$, we can neglect the frictional losses to leading order and only the second term on the right hand side of Eq.\ (\ref{worklambda}) contributes to the work performed against the load, $W \simeq - Q_{\rm tr,ac} \Delta T/T$. If the heat transfer across the system is dominated by the geometric contribution, one finds $Q_{\rm tr} \simeq Q_{\rm tr, ac}$,
and hence that the efficiency approaches $\eta^{\mathrm (he)} \simeq \Delta T/T$. Remarkably, this is just the Carnot efficiency. We thus find that a purely geometric quantum thermal machine reaches the optimal efficiency, and it is the nongeometric contributions to $W$ and $Q_{\rm tr}$ (in the sense of Berry) which are responsible for deviations from the Carnot efficiency. Indeed, 
a finite heat conductance diminishes the efficiency of the heat engine, as do frictional losses described by the first term on the right hand side of Eq.\ (\ref{worklambda}). Note that the contribution of the heat conductance to the transferred heat is proportional to the period of the cycle. This implies that this term is less detrimental to the efficiency as the frequency at which the machine operates increases.
Conversely, by increasing the frequency, the effect of the frictional losses becomes larger.

While the overall efficiency is fundamentally limited to the Carnot limit, there is no fundamental limit to reducing the detrimental effects of the nongeometric contributions. While the frictional forces become arbitrarily small as one approaches the truly adiabatic limit, the limit of a negligible heat conductance $\Lambda_{N+1,N+1}\simeq 0$ can be realized in a topological quantum pump. In such pumps, the ground state is separated from the excited states by a gap. Consequently, the symmetric contributions to $\Lambda_{\mu,\nu}$ -- including the heat conductance -- are strongly suppressed. 

\subsubsection{Refrigerator}
\label{eff_refrigerator}

A refrigerator uses work $W$ performed on the system to remove heat from a cold to a hot reservoir. Thus, we can define a corresponding efficiency or coefficient of performance (COP) as 
\begin{equation}
 \eta^{\rm (fr)} =\frac{-Q_{\rm tr}}{W}.
\label{effx_fr}
\end{equation}
Again focusing on a time reversal invariant system with constant $\Delta T/T$, this efficiency approaches the Carnot limit $\eta^{\mathrm fr}=T/\Delta T$ for zero heat conductance. The efficiency is reduced by a finite heat conductance since, for a refrigerator, its contribution to the numerator has the opposite sign compared to the pumped heat. 

\subsubsection{Heat pump}
\label{eff_pump}

Of course, the device can also be used as an adiabatic heat pump in the absence of a thermal bias $\Delta T/T$. Heat is transported from left to right or vice versa due to the variation of 
${\vec X}$. According to Eq.\ (\ref{worklambda}), we need to exert work $W$ associated with dissipation, even if there is no temperature bias. We can then define a corresponding efficiency of heat pumping through 
\begin{equation}
\label{nobiascop}
\eta^{\rm (pump)} = \frac{|Q_{\rm tr, ac}|}{W} .
\end{equation}
The denominator in this expression is proportional to $\Omega$, so that the efficiency of the heat pump grows as it becomes more adiabatic. 

\section{Examples}
\label{Examples}
We now illustrate the general formalism introduced in the previous sections by two driven systems coupled to thermals baths. One example is referred to as a 
  {\em driven qubit} and consists of a generic two-level system with time-dependent  energies and inter-level transition matrix elements,
  coupled to baths of bosonic excitations. This problem will be solved in the limit of weak coupling to the reservoirs. The second example is
  a {\em driven quantum dot}, which consists of a confined structure with two single-electron levels -- one per spin orientation -- driven by a rotating magnetic field. This problem is solved for weak as well as for strong coupling to  
  spin-polarized electron reservoirs. 

\subsection{Driven qubit}
\label{example-qubit}
\begin{figure}[t]
\centering
\includegraphics[width=\columnwidth]{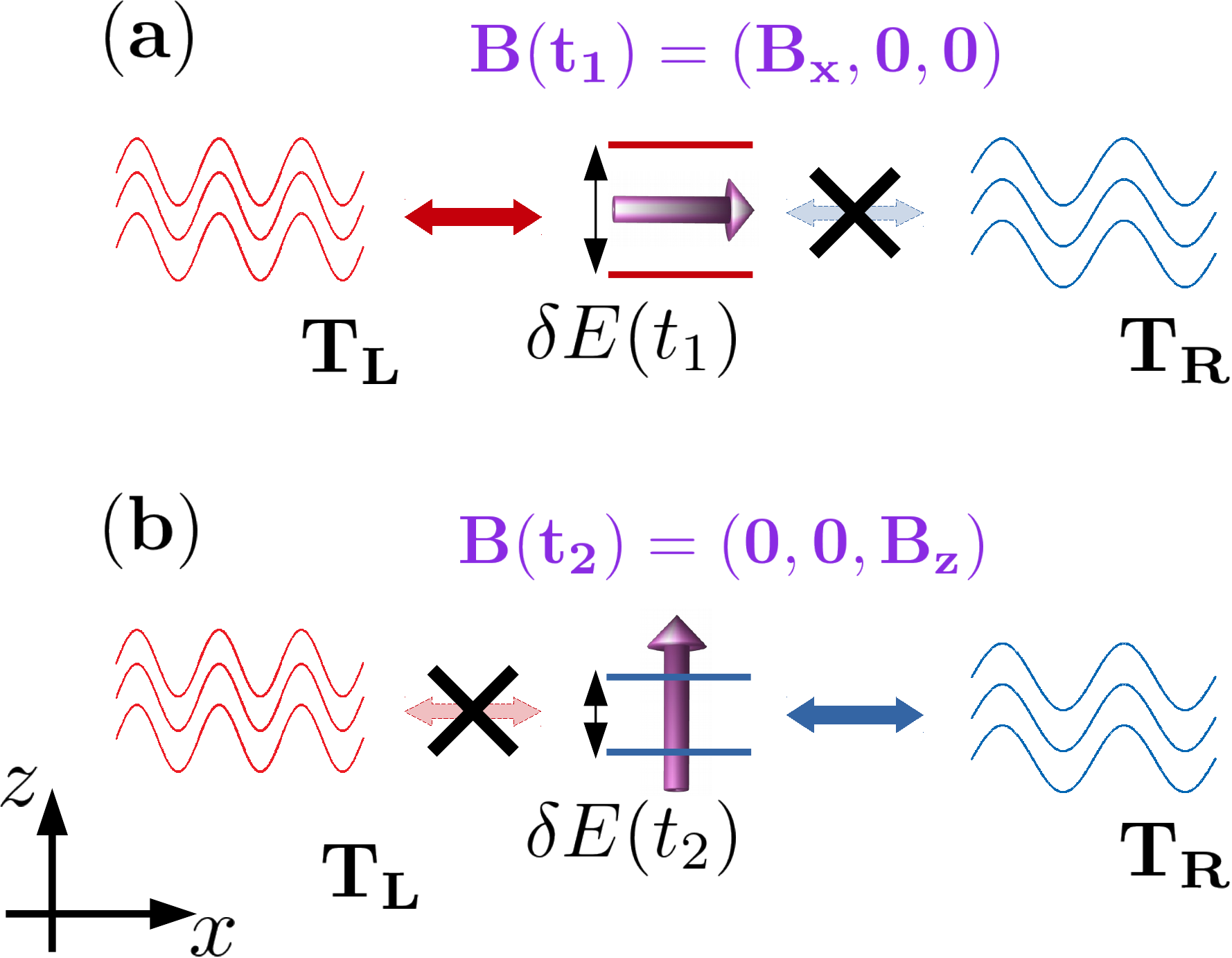}
\caption{Illustration of the q-bit coupled to two bosonic reservoirs by the Hamiltonian of Eq. (\ref{qcont}) with $\hat{\tau}_L=\hat{\sigma}_x$ and $\hat{\tau}_R=\hat{\sigma}_z$, operating as a heat engine.
Panel (a): the q-bit is in one of the states $|x, \pm \rangle$ 
and couples to the reservoir $L$. 
Panel (b): the q-bit is in one of the states $|z, \pm \rangle$ and couples only to the reservoir $R$. The driving changes the energy difference between the two levels. 
}
\label{figure0}
\end{figure}
We consider a generalization of the celebrated spin-boson model, which was introduced in Refs. \cite{sb1,sb2}.
As in those works, we express the Hamiltonian in terms of the Pauli matrices $\hat{\mbox{$\vec{\sigma}$}}=(\hat{\sigma}_x, \hat{\sigma}_y,\hat{\sigma}_z)$ and a magnetic field
 $\vec{B}(t)=\left(B_x(t), B_y(t), B_{z}(t) \right)$. In our case, the latter varies periodically in time.
 The ensuing Hamiltonian reads
\begin{equation}\label{qs}
{\cal H}_S(t)= \vec{B}(t) \cdot \hat{\mbox{$\vec{\sigma}$}}.
\end{equation}
 The reservoirs are represented by the Hamiltonians
\begin{equation}\label{qres}
{\cal H}_{\alpha}=\sum_{k}\varepsilon_{k\alpha}b_{k\alpha}^\dagger b_{k\alpha}, \;\;\; \alpha=L, R,
\end{equation}
with $b_{k\alpha}$ and $b_{k\alpha}^\dagger$ being the annihilation and creation operators of a bosonic excitation. 

The coupling is described by the Hamiltonian ${\cal H}_c={\cal H}_{c,L}+{\cal H}_{c,R}$.
Our generalization with respect to previous works is to consider different types of couplings to the $L$ and $R$ reservoirs. This is motivated by 
the fact that spatial inversion symmetry has to be broken in order to obtain pumping,
as mentioned in Section~\ref{ener-bal}. Concretely, the Hamiltonians read
\begin{equation}\label{qcont}
{\cal H}_{c,\alpha}=\sum_{k}V_{k\alpha}\hat{\tau}_{\alpha}\Big(b_{k\alpha}+b_{k\alpha}^\dagger\Big),
\end{equation}
with  $\hat{\tau}_{L}=\hat{\sigma}_{x}$ and $\hat{\tau}_{R}=\hat{\sigma}_{z}$. Hence, the q-bit couples to the $L$ or $R$ reservoir if it is in a state with a non-vanishing projection 
on the eigenstates $|x,\pm \rangle$ of $\hat{\sigma}_x$ or $|z, \pm \rangle$ of $\hat{\sigma}_z$, respectively.
Any other combination of two Pauli matrices with $\hat{\tau}_{L} \neq \hat{\tau}_{R}$ would also be appropriate, as we will discuss in Section IV.A.3.
Previous works related to heat engines based on q-bits considered the same type of coupling to the two reservoirs and non-adiabatic driving \cite{rob1,rob2,gab,segal2005,nit,hanggi,paula,caru,
karimi,liu2017,wang2018,yamamoto2018,newman} .

The Hamiltonian for the system of Eq. (\ref{qs}) can be transformed to the  basis of instantaneous eigenstates $|j \rangle$, such that ${\cal H}_S(t) |j \rangle= E_j(t) |j \rangle$, $j=1,2$, 
with $E_{1,2}(t)=\mp |\vec{B}|$. 
The resulting transformed Hamiltonian reads $\tilde{\cal H}_{ S}(t)=\hat{U}^{-1}(t) {\cal H}_S(t) \hat{U}(t)$ with $\hat{U}(t)$ being a unitary transformation and 
\begin{equation}\label{hs1}
\tilde{\cal H}_{ S}(t)=E_{1}(t) |1\rangle \langle 1| + E_{2}(t) |2\rangle \langle 2|, 
\end{equation}
Accordingly, the contact Hamiltonian can be also expressed in this basis as
\begin{equation}\label{qcontch}
\tilde{\cal H}_{c,\alpha}(t)=\sum_{k}\sum_{ij}V_{k\alpha}{v}_{\alpha,ij}(t)\hat{\rho}_{ij}(t)\Big(b_{k\alpha}+b_{k\alpha}^\dagger\Big),
\end{equation}
with ${v}_{\alpha,ij}(t)= \left[\hat{U}^{-1}(t) \hat{\tau}_{\alpha}\hat{U}(t)\right]_{ij}$, $\hat{U}(t)$ being the unitary transformation which diagonalizes the Hamiltonian~(\ref{qs}), and $\hat{\rho}_{ij}=|i\rangle \langle j |$.

 Before proceeding to explicit calculations, we can gather some intuition on how the driven q-bit may work as a thermal machine
by using the sketch of Fig.~\ref{figure0}. As a consequence of the driving, the energy of the two levels as well as the coupling to the $L$ and $R$ reservoirs
 change in time according to Eqs. (\ref{hs1}) and  (\ref{qcontch}), respectively. 
Panel (a) represents a situation where the q-bit  at a given time $t_1$ is in one of the  eigenstates of $\hat{\sigma}_x$, hence, it couples 
to the $L$ reservoir and it is completely decoupled from $R$. Panel (b) illustrates the situation where the q-bit is in an eigenstate of $\hat{\sigma}_z$ at a different time $t_2$, 
  therefore  it is coupled to $R$ and decoupled from $L$. In an evolution  from $t_1$ to $t_2$ the energy difference $\delta E(t)=E_2(t)-E_1(t)$ changes. A cycle can be realized when the protocol returns the q-bit to the state 
  of the step (a).  
The paradigmatic Otto cycle corresponds to the extreme situation, where  the q-bit is allowed to thermalize with  $L$ at the step (a) and with $R$ 
 at the step (b), while it evolves decoupled from the two reservoirs at intermediate times
 \cite{karimi,camp}. 
 For the case of adiabatic driving, the changes take place smoothly and the q-bit is coupled  to the two reservoirs at all times. For suitable protocols, the setup may anyway operate as a heat engine or refrigerator, as well as a heat pump. 
 
 We will analyze in detail protocols with two time-dependent parameters of the form $\vec{B}(t)=\left(B_x(t),0, B_z(t)\right)$, with 
\begin{eqnarray}
B_x(t)&=&B_{x,0}+B_{x,1}\cos (\Omega t+\phi), \nonumber \\
 B_z(t) &=&B_{z,0}+B_{z,1}\cos(\Omega t).
 \end{eqnarray}
 These two components of $\vec{B}(t)$ are identified with the time-dependent parameters of Eq. (\ref{hamtot}) as follows 
 \begin{equation}\label{xb}
 \vec{ X}(t) =\left(X_1(t),X_2(t)\right) \equiv  \left(B_z(t),B_x(t) \right).
 \end{equation}
 In addition, we will consider a constant difference of temperature $\Delta T$, which defines $\dot{X}_3=\Delta T/T$.
We will solve the problem in the limit of very weak coupling between the qubit and the reservoirs (small $V_{k\alpha} $).

\subsubsection{Master equation approach}
We follow the procedure of Refs.~\onlinecite{janine,janine1,janine2}, which consists in solving the time-dependent master equation by performing an adiabatic expansion along the
 lines  of the general formalism of Section~\ref{linear_ad}.
The basic idea is to describe the evolution of the  population probabilities of the eigenstates of $\tilde{\cal H}_{ S}(t)$, represented by the vector $\mathbf{p}(t)= \left( p_1(t), p_2(t) \right)$,  in terms of 
a master equation where the effect of the coupling to the reservoirs is treated at the lowest order of perturbation theory (first order in $|V_{k\alpha}|^2$). The master equation reads, 
\begin{equation}
\label{meex1}
\frac{d}{dt}\mathbf{p}(t)=\sum_\alpha\mathbf{M}_\alpha(\vec{B})\cdot \mathbf{p}(t),
\end{equation}
where $\mathbf{M}_\alpha(\vec{B})$ is a $\text{2}\times \text{2}$ matrix representing the instantaneous transition rates corresponding to the reservoir $\alpha$, which is  given by
\begin{align}
\mathbf{M}_\alpha(\vec{B})=\begin{bmatrix}
-\Gamma_{1\rightarrow 2}^\alpha(\vec{B}) & \Gamma_{2 \rightarrow 1}^\alpha(\vec{B}) \\
\Gamma_{1\rightarrow 2}^\alpha(\vec{B}) & -\Gamma_{2\rightarrow 1}^\alpha(\vec{B})
\end{bmatrix}.
\label{tunmatr}
\end{align} 
Here we stress that the instantaneous rates depend on time through the parameters $\vec{B}$,  as indicated in Eq. (\ref{xb}).
We have introduced  the following definitions
\begin{align}
&\Gamma_{1\rightarrow 2}^\alpha(\vec{B})=\lambda_\alpha(\vec{B}) \Big[\gamma_\alpha \Big(\delta E \left(\vec{B}\right)\Big)+\tilde{\gamma}_\alpha\Big(-\delta E\left(\vec{B}\right)\Big)\Big],\nonumber \\
&\Gamma_{2\rightarrow 1}^\alpha(\vec{B})=\lambda_\alpha(\vec{B}) \Big[\tilde{\gamma}_\alpha \Big(\delta E(\vec{B}) \Big)+{\gamma}_\alpha \Big(-\delta E(\vec{B}) \Big)\Big],
\label{repre}
\end{align}
with 
\begin{eqnarray}
\gamma_{\alpha}(\varepsilon) &= &n_{\alpha}(\varepsilon)\Gamma_{\alpha}(\varepsilon)/\hbar, \nonumber \\
\tilde{\gamma}_{\alpha}(\varepsilon) &=& [1+n_{\alpha}(\varepsilon)]\Gamma_{\alpha}(\varepsilon)/\hbar,
\end{eqnarray}
while
 $\delta E(\vec{B})=E_2(\vec{B})-E_1(\vec{B})$ and $\lambda_\alpha(\vec{B})=v_{\alpha,12}(\vec{B})v_{\alpha,21}(\vec{B})$. For $\hat{\tau}_L=\hat{\sigma}_x$ and $\hat{\tau}_R=\hat{\sigma}_z$ we have
\begin{equation}
\lambda_L(\vec{B})=\frac{B_x^2(t)}{B_z^2(t)+B_x^2(t)}, \;\;\;\;\;\;\lambda_R(\vec{B})=\frac{B_z^2(t)}{B_z^2(t)+B_x^2(t)}.
\label{lambda1}
\end{equation} 
$n_{\alpha}(\varepsilon)$ is  the Bose-Einstein distribution for bath $\alpha$ and $\Gamma_{\alpha}(\epsilon)$ is the corresponding spectral density,
which we assume to be Ohmic
\begin{equation}\label{ombath}
\Gamma_{\alpha}(\epsilon)=\Gamma_\alpha\, \epsilon\, e^{-\epsilon/\epsilon_{\rm C}}, \;\;\;\text{with }\epsilon >0,
\end{equation}
$\epsilon_{\rm C}$ being the cut-off frequency. Since, according to Eq.~(\ref{ombath}), there are no negative-energy states in the bath, we set $\gamma_\alpha [-\delta E(\vec{B})] = 
\tilde{\gamma}_\alpha [-\delta E(\vec{B}) ]=0$ (notice that  $\delta E(\vec{B})$ is positive by definition).

Following Refs.~\onlinecite{janine1,janine2},
the population can be expanded in different orders of the driving frequency $\Omega$. 
Here we keep only the zeroth-order (instantaneous) term $\mathbf{p}^{({\rm i})}$, and first-order (adiabatic) term $\mathbf{p}^{({\rm a})}$ such that
\begin{equation}
\mathbf{p}(t)=\mathbf{p}^{({\rm i})}(t)+\mathbf{p}^{({\rm a})}(t).
\label{probabil}
\end{equation}
The solution of the master equation~(\ref{meex1}) order by order in $\Omega$, leads to
\begin{align}
\sum_\alpha\mathbf{M}_\alpha(\vec{B})\cdot\mathbf{p}^{({\rm i})}(t)=0,
\label{insad1}
\end{align}
and
\begin{align}
\frac{d}{dt}\mathbf{p}^{({\rm i})}(t)=\sum_\alpha\mathbf{M}_\alpha(\vec{B})\cdot\mathbf{p}^{({\rm a})}(t).
\label{insad2}
\end{align}
The adiabatic correction can be written in terms of instantaneous contributions as
\begin{equation}
\mathbf{p}^{({\rm a})}(t)=\sum_\alpha\left[\bar{\mathbf{M}}_\alpha(\vec{B})\right]^{-1}\cdot\frac{d}{dt}\mathbf{p}^{({\rm i})}(t),
\label{papi}
\end{equation}
where the matrix $\left[\bar{\mathbf{M}}_\alpha(\vec{B})\right]^{-1}$ includes the normalization condition for the adiabatic probabilities~\cite{janine}.
We obtain two additional equations from the conservation of the probability, namely $\sum_{j}p_{j}^{({\rm i})}(t)=1$ and $\sum_{j}p_{j}^{({\rm a})}(t)=0$.

The instantaneous (${\rm i}$), adiabatic (${\rm a}$) and thermal (${\rm th}$) contributions to the heat current flowing in reservoir $\alpha$ as functions of time are given by
\begin{eqnarray}
J_{\alpha}^{({\rm i/a})}(t) &= &\delta E(\vec{B}) \left[\mathbf{M}_\alpha(\vec{B})\cdot \mathbf{p}^{\rm (i/a)}(t)\right]_{11},\nonumber \\
J_{\alpha}^{({\rm th})}(t) & = &\delta E(\vec{B}) \left[\mathbf{M}_\alpha(\vec{B})\cdot \mathbf{p}^{\rm (i)}_{\Delta T}(t)\right]_{11},
\label{asymJ}
\end{eqnarray}
where $\mathbf{p}_{\Delta T}^{({\rm i})}$ is the instantaneous probability vector in the presence of the thermal bias $\Delta T$.
We can now calculate the different linear-response components of the heat current defined in Eq.~(\ref{therpum}) as follows
\begin{equation}
J_{{\rm tr,ac}/\Delta T}^{Q}  =  \frac{\Omega}{2 \pi} \int_0^{2\pi/\Omega} dt \; J_{\rm R}^{({\rm a/th})}(t), 
\label{J}
\end{equation}
while the instantaneous component vanishes when averaged over the period. 

On the other hand, the net work  developed by the ac forces, corresponding to Eq. (\ref{worklambda}) can also be calculated in the master equation approach.
To this end we write the  total energy of the qubit at a particular time $t$ as
\begin{equation}
{E}_{\rm tot}(t)=E_1(t) p_1(t) + E_2(t) p_2(t),
\end{equation}
where the probabilities are given by the sum of the instantaneous $p_j^{\rm (i)}$, the adiabatic $p_j^{\rm (a)}$ and thermal $p_j^{\rm (th)}$ components.
The time derivative of the total energy contains two contributions,
\begin{eqnarray}
\frac{d{E}_{\rm tot}}{dt}&=&\sum_{j=1}^2 \left(  \frac{dE_j(t)}{dt} p_j(t) +
 E_j(t) \frac{dp_j(t)}{dt} \right)
\end{eqnarray}
These are the power delivered by the ac sources
\begin{equation}
 {P}(t)=\frac{dE_1(t)}{dt} p_1(t) + \frac{dE_2(t)}{dt} p_2(t),
\end{equation}
and the heat temporarily stored in the q-bit. Thus, the total work over a cycle reads
\begin{align}\label{wqub}
W=\int_0^{2\pi/\Omega} dt \Big(\frac{dE_1}{dt}p_1(t)+\frac{dE_2}{dt}p_2(t)\Big),
\end{align}
where both instantaneous, adiabatic, and thermal components of the probabilities $\mathbf{p}(t)$ contribute.
The contribution due to the instantaneous components represents the work done by the conservative forces, while the other terms will contribute to the non-conservative work defined in Eq. (\ref{worklambda}). The explicit expressions for the different components of $\mathbf{p}(t)$ for the driving protocol of Eq. (\ref{xb}) are presented in 
Appendix~\ref{linearr}. We notice that the terms originating from the coupling Hamiltonian in Eq.~(\ref{qcontch}), could in principle contribute to $W$ and  can be calculated from the time average of    $\langle \dot{\tilde{{\cal H}}}_{c,\alpha} \rangle$. However, this term is neglected in the limit of very small $V_{k\alpha}$. In fact, its contribution to the work per cycle is smaller (by at least a factor of $|V_{k\alpha}|$) than the contribution to the work due to $\tilde{H}_S(t)$.

\subsubsection{Geometrical properties}
\label{thadex1}
We now derive the expressions corresponding to Eqs. (\ref{pumpedheatlambda2}) and  (\ref{worklambda}) within the formalism of the master equation. These can be derived 
from Eqs. (\ref{J}) and (\ref{wqub}). We get
\begin{eqnarray}
Q_{\rm tr,ac} &=& \int_0^{2\pi/\Omega} dt \; \mathbf{M}_R^{(h)}(\vec{B})\cdot\mathbf{p}^{({\rm a})}(t),\label{currheatad} \\
W &=&\int_0^{2\pi/\Omega} dt \; \frac{d\mathbf{E}}{dt} \cdot \left[\mathbf{p}^{({\rm a})}(t)+ \mathbf{p}^{({\rm i})}_{\Delta T}(t) \right],
\label{currpowad}
\end{eqnarray}
where
\begin{align}
\mathbf{M}_R^{(h)}(\vec{B})=
\delta E(\vec{B}) \begin{bmatrix}
 -\Gamma_{1\rightarrow 2}^R(\vec{B})
 \\
\Gamma_{2\rightarrow 1}^R(\vec{B})
\end{bmatrix}^{T}.
\end{align}
and ${\bf E}((\vec{B}))=\left(E_1(t),\; E_2(t) \right)$.
Using Eq. (\ref{papi}) and
\begin{equation}\label{dpi}
\frac{d {\bf p}^{(i)}}{dt}= \sum_{\ell=1}^2 \frac{\partial {\bf p}^{(i)}}{\partial B_{\ell}}  \dot{B}_{\ell}
\end{equation}
the pumped heat given by Eq. (\ref{currheatad}) can be written as in Eq. (\ref{pumpedheatlambda2}), by identifying
\begin{equation}
\Lambda_{3,\ell}(\vec{B})=\mathbf{M}^{(h)}_R(\vec{B})\cdot\bar{\mathbf{M}}^{-1}(\vec{  B}) \cdot\frac{\partial{\mathbf{p}}^{({\rm i})}}{\partial B_{\ell}},\;\;\; \ell=1,2.
\end{equation}
In the present configuration, the explicit calculation of these coefficients show that $\Lambda_{3,\ell}= -\Lambda_{\ell,3}$, up to a function that vanishes upon integrating over the period.
This means that these terms are components of  the antisymmetric thermal tensor
$\Lambda^A_{\mu,\nu}$.  
The other components of the tensor can be derived from the first terms ($\propto \mathbf{p}^{({\rm a})}(t)$) of Eq. (\ref{currpowad}).
More precisely, using Eq. (\ref{papi}) with Eq. (\ref{dpi}), and expressing
\begin{eqnarray}\label{el}
\frac{d {\bf E}}{dt}= \sum_{\ell=1}^2 \frac{\partial {\bf E}}{\partial B_{\ell}} \cdot \dot{B}_{\ell},
\end{eqnarray}
we find
\begin{equation}
\Lambda_{\ell,\ell^{\prime}}(\vec{B})= \frac{\partial {\bf E}}{\partial B_{\ell}} \cdot \bar{\mathbf{M}}^{-1}(\vec{B}) \cdot \frac{d {\bf p}^{(i)}}{\partial B_{\ell^{\prime}}} ,\;\;\; \ell,\ell^{\prime}=1,2.
\label{lambdallp}
\end{equation}
We can see that these terms satisfy $\Lambda_{\ell,\ell^{\prime}}=\Lambda_{\ell^{\prime},\ell}$, as explicitly shown in Eq. (\ref{aplamqbit}).
Hence they are components of the symmetric tensor $\Lambda^S_{\mu,\nu}$.

 On the other hand, by using the fact that we can define a relation of the form $\dot{B}_{\ell}= g_{\ell}(\vec{B}) \Omega $ for the protocol of Eq. (\ref{xb}), we can express the total work in terms of purely geometric quantities, by rewritting Eqs. (\ref{currheatad}) and  (\ref{currpowad}) in terms of the  vector potentials of Eqs. (\ref{amu}) and
 (\ref{atildemu}). In the present case, they read
 \begin{eqnarray}\label{aas}
 \vec{A}^A_3(\vec{B}) &=& \left( \Lambda_{3,1}^A(\vec{B}),\Lambda_{3,2}^A(\vec{B}) \right),\nonumber \\
 \vec{A}^S_{\ell}(\vec{B}) &=& \left( \Lambda_{\ell,1}^S(\vec{B}),\Lambda_{\ell,2}^S(\vec{B}) \right), \;\;\; \ell =1,2,\nonumber \\
   \tilde{\vec{A}}(\vec{B}) &= & \Omega \sum_{\ell=1}^2 g_{\ell}(\vec{B}) \left(\Lambda_{\ell,1}^S(\vec{B}), \Lambda_{\ell,2}^S(\vec{B}) \right),
 \end{eqnarray}
 We have highlighted the antisymmetric and symmetric character in each case. Notice that, according to the analysis of Sections \ref{ener-bal} and \ref{geoF}, the symmetric component contributes purely to dissipation of energy and entropy production, while the antisymmetric one is related to useful work. 
 
  In order to characterize the performance of the heat engine and refrigerator  as in Eqs. (\ref{effx_he}) and (\ref{effx_fr}) we also need the heat transported in one period as a response to the thermal bias. It reads
  \begin{equation}
 Q_{\rm tr, \Delta T}=  \int_0^{2\pi/\Omega} dt \; J_{\rm R}^{({\rm th})}(t)
 \end{equation}
 with $J_{\rm R}^{({\rm th})}(t)$ defined in Eq. (\ref{asymJ}). This component is not geometric and we recall that 
 the total transported heat is $Q_{\rm tr}=Q_{\rm tr, ac} + Q_{\rm tr, \Delta T}$.

 According to our conventions, the contribution to the contour integral of the first component of Eq. (\ref{w-a}) is always positive and is the portion related to the net dissipated power and entropy production due to the ac driving. Instead,  the second one, also defining $Q_{\rm tr, ac}$ in Eq. (\ref{q-a}), 
 can have any sign. In the case of a heat engine, $Q_{\rm tr, ac}$ and $Q_{\rm tr, \Delta T}$ have the same sign, i.e. the pumped heat flows in the same direction as the component induced by the temperature bias.
 As a consequence, it generates useful work that  can be absorbed by the ac sources. Notice that in such a case,  the second term of Eq. (\ref{w-a}) has an opposite sign to the first.
 In the refrigerator, it is the opposite. Irrespectively of the sign of $Q_{\rm tr, ac}$, which determines that the system operates as a heat engine or a refrigerator, the crucial quantity to optimize is the integral of $\vec{A}^A(\vec{B}) $ over a suitable chosen closed path in the parameter space.

\subsubsection{Results}
\label{reultsqbt}
We present some results for specific parameters of the driving protocol defined in Eq. (\ref{xb}).

We start by analyzing the case with $\Delta T=0$ and showing that a necessary condition for the heat currents to be finite is that the coupling to the left and right reservoirs are different, i.~e. $\hat{\tau}_L\neq \hat{\tau}_R$. In fact,
let us notice that these couplings determine the functions $\lambda_L(\vec{B})$ and $\lambda_R(\vec{B})$. If we assume symmetric couplings, we have $\lambda_L(\vec{B})=\lambda_R(\vec{B})$ and $\Gamma_L=\Gamma_R$. Therefore, we get  ${\bf M}_L(\vec{B})={\bf M}_R(\vec{B})$ in Eq. (\ref{tunmatr}). After replacing the latter matrices
 in Eq.~(\ref{asymJ}), we get
  $J^{(a)}_L(t)=J^{(a)}_R(t)$ 
at every time.
This implies that the currents obtained by averaging over one period, i.~e. $J^{Q}_{\rm tr,L}$ and $J^{Q}_{\rm tr,R}\equiv J^{Q}_{\rm tr,ac}$, must be equal to zero in order to agree with Eq.~(\ref{net}). 
Interestingly, one can check by means of the explicit calculations that the adiabatically pumped current in one period $J^{Q}_{\rm tr,ac}$ is zero even if one allows $\Gamma_L$ and $\Gamma_R$ to be different.
Moreover, we verified that the magnitude of the pumped heat current depends on the chosen combinations of Pauli matrices  (see Appendix ~\ref{linearr}).
The maximum pumping for the protocol of Eq. (\ref{xb}) corresponds to 
 $\mathcal{H}_{\rm c,\alpha}$ containing $\hat{\tau}_L=\hat{\sigma}_x$ and $\hat{\tau}_R=\hat{\sigma}_z$, as in Eq. (\ref{qcont}).
As a matter of fact, in the other two combinations ($\hat{\tau}_L=\hat{\sigma}_x$, $\hat{\tau}_R=\hat{\sigma}_y$, and $\hat{\tau}_L=\hat{\sigma}_y$, $\hat{\tau}_R=\hat{\sigma}_z$) one obtains half the magnitude.  

\begin{figure}[!htb]
\centering
\includegraphics[width=\columnwidth]{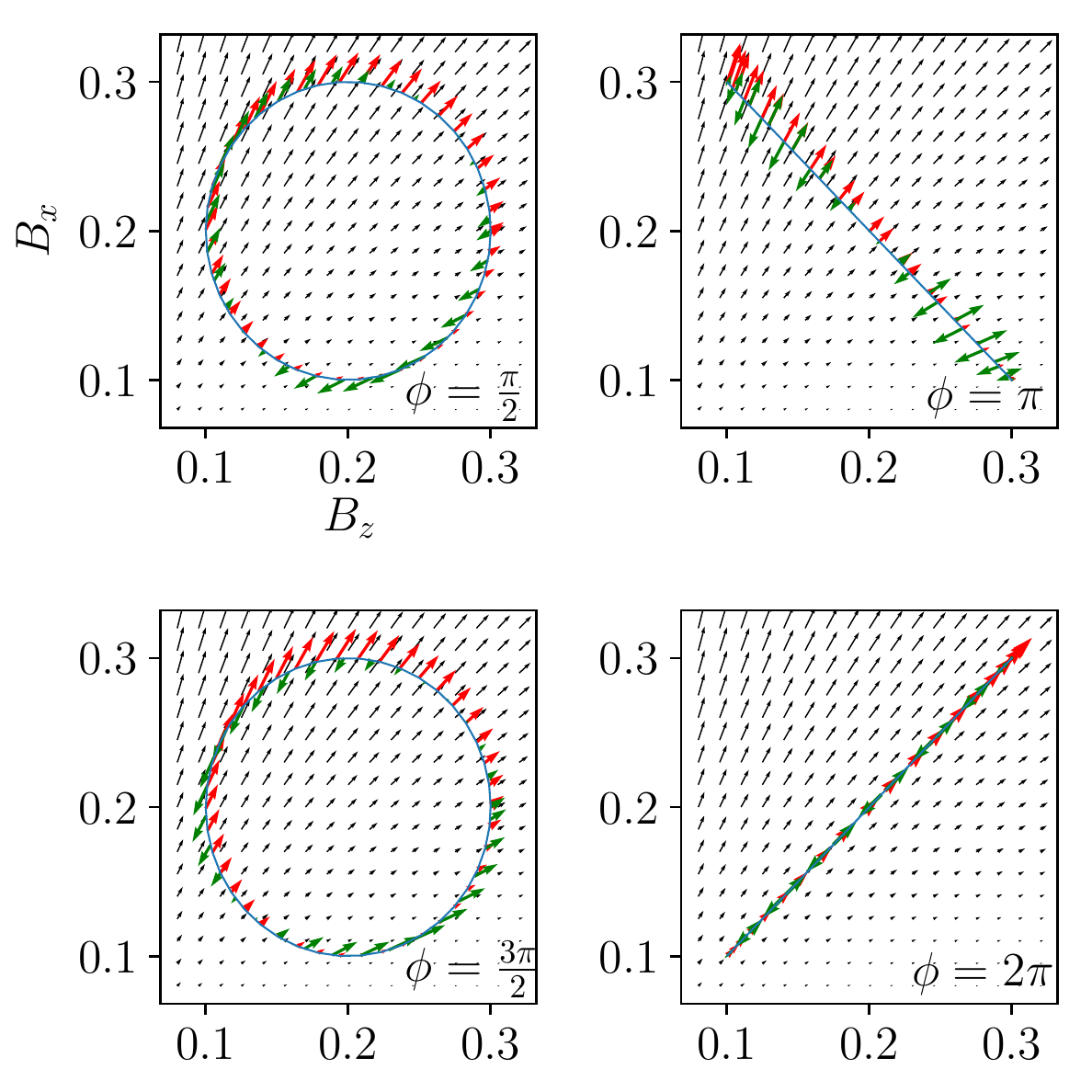}
\caption{Vectors $\vec{A}^{A}$ and $ \tilde\vec{A}^{S}$. Black and red arrows represent the vector $\vec{A}_3^{A}(\vec{B})\equiv \left( \Lambda_{3,1}^A(\vec{B}), \Lambda_{3,2}^A(\vec{B}) \right)$ in the parameter space, while the green arrows represent the vector $ \tilde\vec{A}^{S}$ defined in Eq.~(\ref{atildemu}). The blue line is the closed path corresponding to
  the driving protocol in Eq.~(\ref{xb}) with 
  $B_{x,0}=B_{z,0}=0.2k_{\rm B}T$, $B_{x,1}=B_{z,1}=0.1k_{\rm B}T$. The other parameters are $\Gamma_{\rm L}=\Gamma_R=0.2$ and $\epsilon_{\rm C}=100k_{\rm B}T$ and define the spectral properties of the
  bosonic bath as indicated in Eq. (\ref{ombath}).}
\label{figcont}
\end{figure}

We now turn to analyze the geometric properties, which can be fully characterized 
 by the vector potentials $\vec{A}^{A}(\vec{B})$ and $\tilde{\vec{A}}^{S}(\vec{B})$, entering Eqs. (\ref{w-a}) and (\ref{q-a}). These vectors are represented with arrows in the parameters space in Fig.~\ref{figcont}. In the Fig.~\ref{figcont} we show several paths, which are plotted in blue, corresponding to the protocol of Eq. (\ref{xb}) with different relative phases $\phi$. 
 This provides a visual representation of the magnitude of $Q_{\rm tr, ac}$ and the
two types of geometric components of $W$. 
 In all the cases we 
 represent with red arrows the vector $\vec{A}^A(\vec{B})$ along the path while the green arrows represent the vector potential $\tilde{\vec{A}}^S(\vec{B})$ along the same protocol (note that  $\tilde{\vec{A}}^{S}(\vec{B})$ is inherently associated with the protocol and cannot be defined outside it). The latter vectors follow the circulation of the path. Thus, they lead to a positive non-vanishing contribution to $W$ for all the values of $\phi$. Instead, the vectors $\vec{A}^{A}(\vec{B})$ are in general opposite to the circulation of the path along some pieces. In particular, 
 for trajectories like the ones corresponding to $\phi=n \pi$, they are parallel to the circulation along half of the path and antiparallel in the other half, leading to a vanishing result of the integral. 
 
\begin{figure}[!htb]
	\centering
\includegraphics[width=0.9\columnwidth]{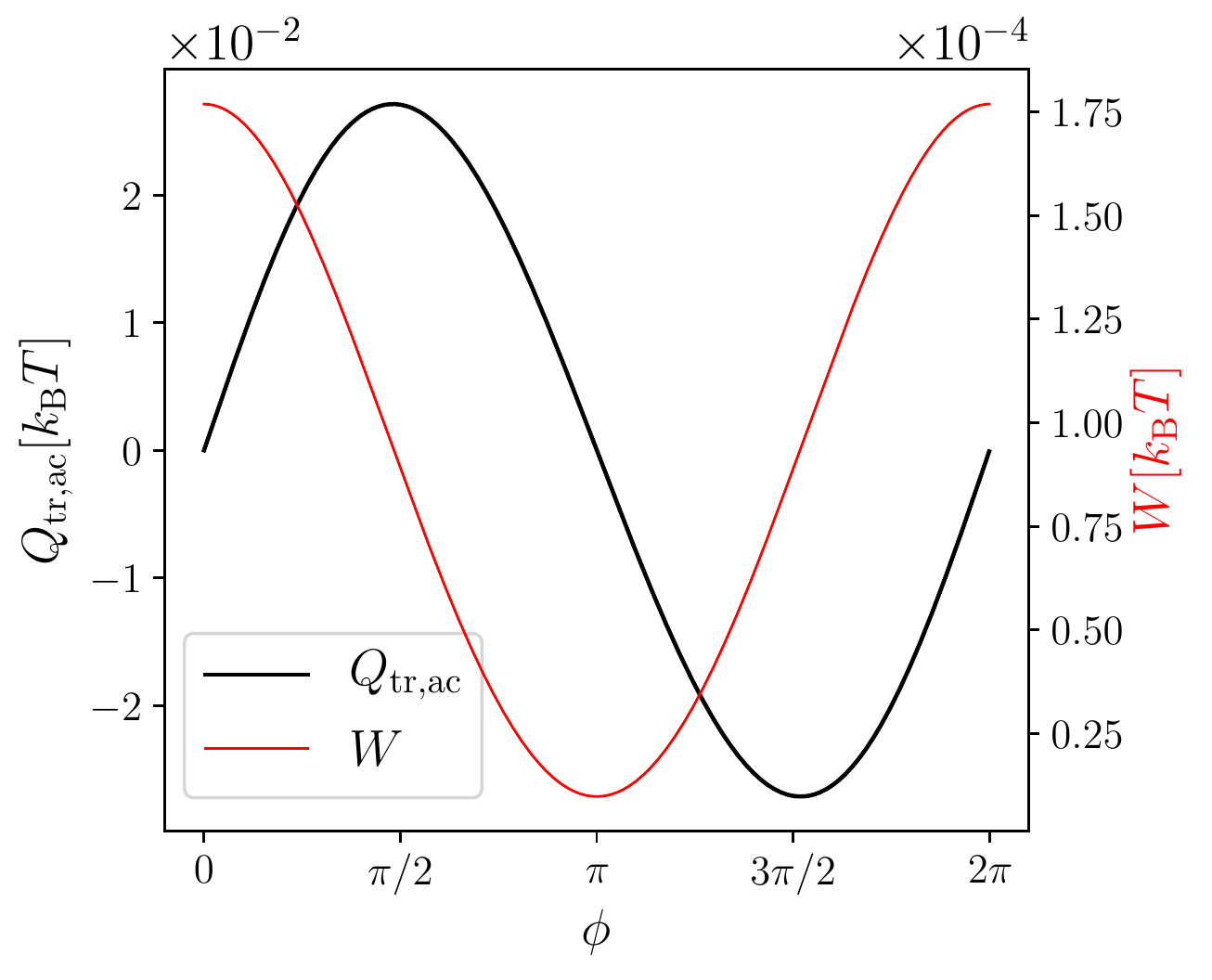}
\caption{Adiabatically pumped heat $Q_{\rm ac}$ and total work $W$ versus the phase lag $\phi$ in the weak pumping limit for $\Delta T=0$. Same parameters as in Fig.~\ref{figcont}.
}
	\label{fig:currlrs}
\end{figure}

In Fig.~\ref{fig:currlrs} we plot the adiabatically pumped heat current $Q_{\rm tr,ac}$, black curve, as a function of the phase lag $\phi$ in the weak pumping limit.
The latter corresponds to considering values of $B_{x,1}$ and $B_{z,1}$ small enough so that $\oint \vec{A}^A_3 \cdot d \vec{B}$ in Eqs.~(\ref{w-a}) and (\ref{q-a}) is proportional to the area, in the parameter space, enclosed by the closed contour defining the protocol. Indeed, using the Green's theorem, these integrals can be written as a surface integral of the derivatives of $\vec{A}^A_3$ with respect to $\vec{B}$. When $B_{x,1}$ and $B_{z,1}$ are small, such derivatives do not depend on $\vec{B}$ and can be factorized outside the integral.
Accordingly, as shown in Fig.~\ref{fig:currlrs}, the pumped heat current (black curve) behaves as a sine function of $\phi$, which vanishes at $\phi=0$.
In particular, we note that a heat current is extracted from the reservoir R when $\phi$ is between 0 and $\pi$ and injected for $\pi < \phi < 2\pi$.
The dependence of the total work $W$ developed by the ac sources with respect to the phase lag $\phi$ is is also plotted in Fig.~\ref{fig:currlrs} (red curve) using the same parameters as for the heat current.
We notice that $W$ is finite in the whole range of values of $\phi$, behaving like a cosine function with a vertical offset, hence, it is non-vanishing in any case.

\begin{figure}[htb]
	\centering
\includegraphics[width=\columnwidth]{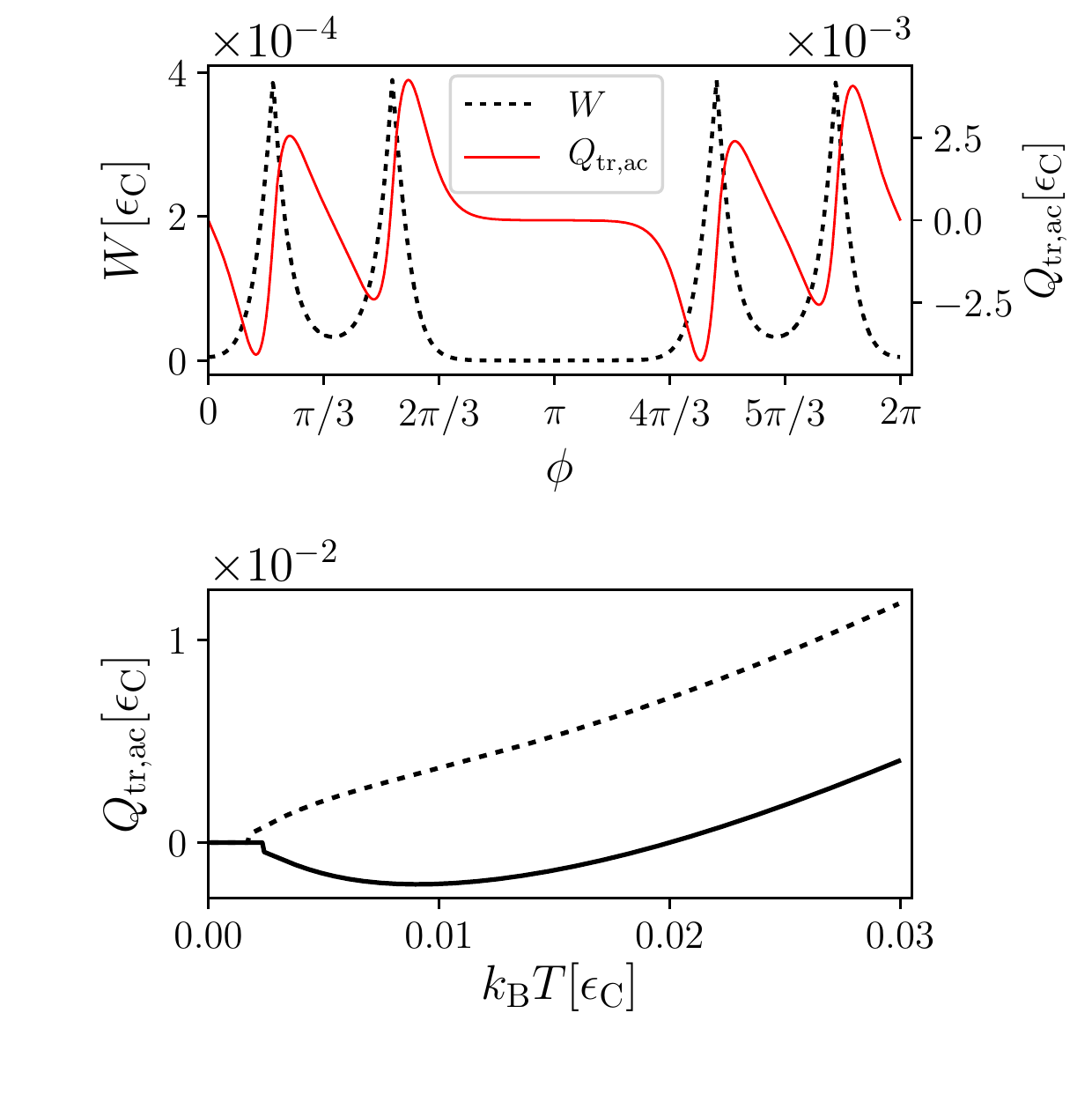}
\caption{Top panel: Pumped heat and work  versus the phase difference between adiabatically-driven system parameters for $k_{\rm B}T=0.01\epsilon_{\rm C}$. Bottom panel: normalized pumped heat currents flowing in the left and right lead for $\phi=\pi/2$. We have used the following parameters: $\Gamma_L=\Gamma_R=1/5$, $B_{z,0}=0.06\epsilon_{\rm C}$ ($B_{z,0}=0.04\epsilon_{\rm C}$ for the dashed lines in the bottom panel), $B_{x,0}=0.03\epsilon_{\rm C}$, $B_{x,1}= B_{z,1}=0.07 \epsilon_{\rm C}$, and $\Delta T=0$.}
	\label{fig:linres}
\end{figure}

\label{srpump}
\begin{figure}[!htb]
	\centering
\includegraphics[width=\columnwidth]{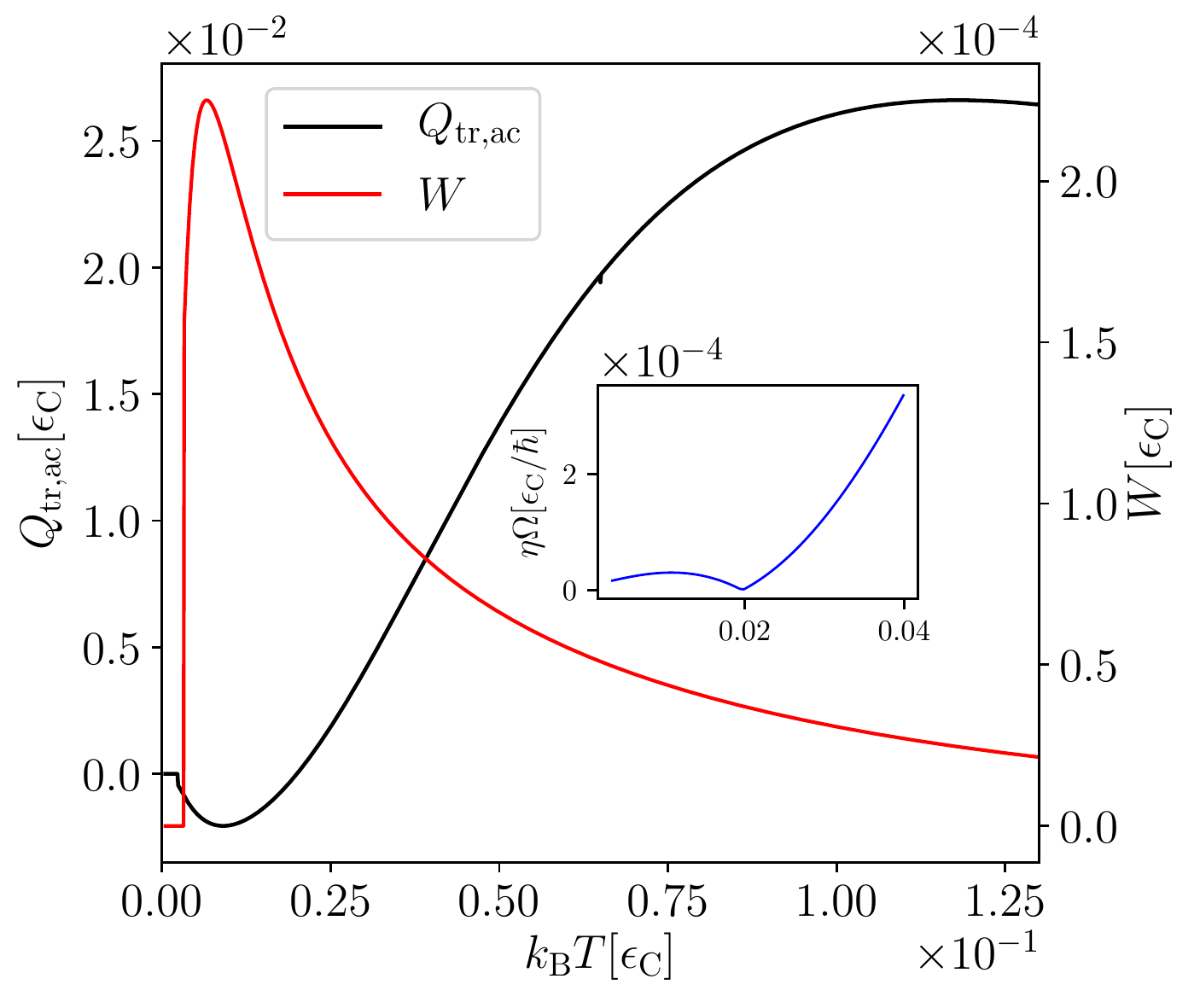}
\caption{Pumped heat and work versus reference temperature $T$. Inset: efficiency of a heat pump for $\Delta T=0$ as a function of $k_{\rm B}T$. Same parameters as in Fig.~\ref{fig:linres} for the solid curves.}
	\label{fig:linresT}
\end{figure}

 In what follows, we show some results for  the strong pumping regime corresponding to larger amplitudes of $B_{x,1}$ and $B_{z,1}$. In the top panel of Fig.~\ref{fig:linres}, we plot the heat pumped and the work performed in a period by the ac source as functions of the phase lag $\phi$. As in the case of weak pumping previously analyzed, 
  the pumped heat as well as the work performed by the ac sources are equal to zero at $\phi=0$ and $\pi$, since the contour has no area (see Fig. \ref{figcont}). For other parameters, it is difficult to make a simple argument to explain in which direction is the heat pumped. In fact, we see that $Q_{\rm tr,ac}$ changes sign many times between $\phi=0$ and $\phi=2\pi$, whereas
$W$ shows multiple positive peaks.
In the bottom panel of Fig.~\ref{fig:linres} we plot the pumped heat in the absence of thermal bias as a function of temperature. For a suitable choice of parameters (relative to the solid curves), the direction of the flow of adiabatic heat can be reversed just by increasing the temperature of the reservoirs.
In Fig.~\ref{fig:linresT} we plot the variation of the heat pumped and the work performed by the ac source, namely $Q_{\rm tr,ac}$ and $W$, as a function of the temperature $T$.
We note that $W$ is always positive, as expected, and is non monotonous (displaying a maximum).
$Q_{\rm tr,ac}$ are the same data as in Fig.~\ref{fig:linres} bottom, but plotted in a larger range of temperatures.
$Q_{\rm tr,ac}$ is non monotonous too and changes sign, going from negative values for small $T$ to positive values at around $k_BT=0.02\epsilon_{\rm C}$.
The inset of Fig.~\ref{fig:linresT} shows the efficiency $\eta^{\rm (pump)}$, defined in Eq.~(\ref{nobiascop}), of the system operated as a heat pump as a function of $T$.
The non-monotonic behavior simply reflects the fact that, in the strong pumping regime, the heat currents change sign at around $k_BT=0.02\epsilon_{\rm C}$, as shown in Fig.~\ref{fig:linres}.

\begin{figure}[!htb]
	\centering	\includegraphics[width=0.9\columnwidth]{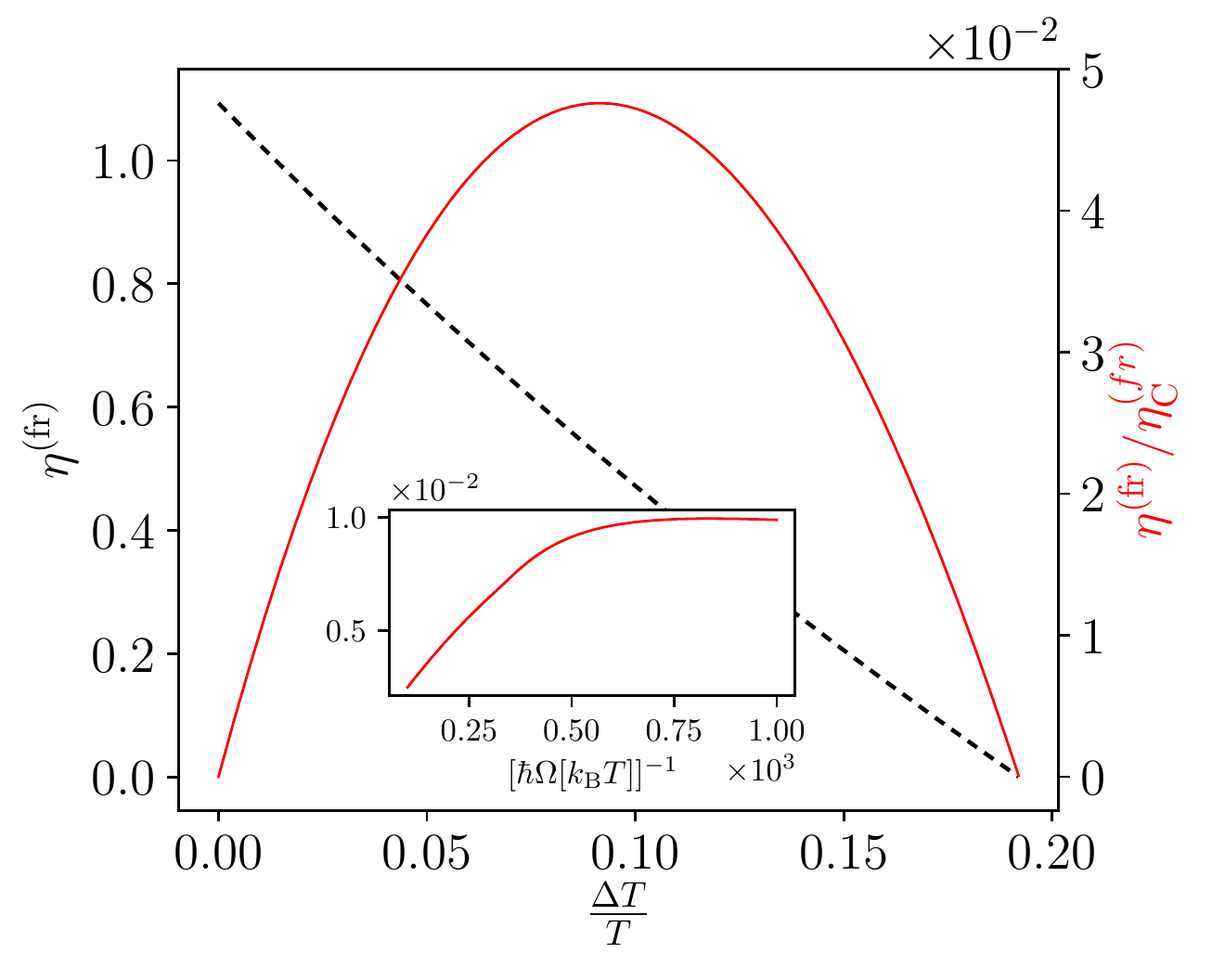}
\caption{Coefficient of performance for refrigeration (black dashed curve for absolute value and red curve for normalized to the Carnot value) versus $\Delta T$, for $\hbar \Omega=
k_{\rm B}T/100$ and versus $\Omega$ (in the inset), for $\Delta T=T/500 $. We use the following parameters: $\Gamma_{\rm L}=\Gamma_{\rm R}=0.2$, $B_{z,1}=10k_{\rm B}T$, $B_{x,0}=20k_{\rm B}T$, $B_{x,1} = 30k_{\rm B}T$, $B_{z,0}=7 k_{\rm B}T$, $\epsilon_{\rm C}=120 k_{\rm B}T$, $\phi=\pi/2$.}
\label{fig:heatpmpcopqbtdt2}
\end{figure}

Finally, in Fig.~\ref{fig:heatpmpcopqbtdt2} we assess the performance of the driven q-bit as a refrigerator which removes heat from the cold reservoir ($R$) even in the presence of a positive thermal bias $\Delta T$, i.~e. for $T_R<T_L$. Given this temperature bias, we focus on a protocol with $\phi=\pi/2$ and the same driving parameters as in Fig. \ref{fig:currlrs}, in which case, we already know from the analysis of this figure, 
that heat is pumped from the coldest reservoir and the heat current at zero bias is maximum. 

We plot the COP $\eta^{\rm(fr)}$ as a black dashed curve, defined in Eq.~(\ref{effx_fr}), and the normalized COP $\eta^{\rm(fr)}/\eta_{\rm C}^{\rm (fr)}$ (red curve) as functions of $\Delta T$, where $\eta^{\rm (fr)}_C = T/ \Delta T$ is the Carnot COP.
Starting from $\Delta T=0$, where $\eta^{\rm(fr)}$ is roughly equal to 1.1, the plot shows that $\eta^{\rm(fr)}$ monotonously decreases with $\Delta T$.
This behavior can be understood by recalling that the refrigeration mode results from a competition between the  heat induced by the temperature difference and the pumped heat
against the thermal bias. In fact,
$Q_{\rm tr}$ is made up of two components: i) the component $Q_{ \Delta T}=2 \pi J_{{\rm tr},\Delta T}^Q/\Omega $, which is the heat current  flowing from the hot to the cold reservoir during one period, therefore entering the reservoir $R$ ($Q_{\Delta T}>0$). This component increases linearly with $\Delta T$; ii)  $Q_{\rm tr, ac}$, which is the pumped heat current extracted from the cold reservoir $R$ ($Q_{\rm tr, ac}<0$), which is independent of $\Delta T$. Therefore $Q_{\rm tr}$ remains negative as long as $Q_{ \Delta T}$ is not large enough to compensate $Q_{\rm tr, ac}$.
This occurs at $\Delta T\simeq 0.19 \;T$, where the total transported heat $Q_{\rm tr}$ vanishes, i.~e. the thermal machine is no longer a refrigerator (a further increase of $\Delta T$ leads to a sign reversal of the heat current).

On the other hand, the ratio $\eta^{\rm(fr)}/\eta_{\rm C}^{\rm (fr)}$ (red curve) is bell-shaped, since this ratio becomes $\propto \Delta T$.  
In the inset of Fig.~\ref{fig:heatpmpcopqbtdt2} we plot the normalized COP as a function of the inverse of the driving frequency $\Omega$. Since $Q_{ \Delta T} \propto \Omega^{-1}$,
increasing the frequency -- within the adiabatic regime -- favors the pumping component $Q_{\rm ac}$  relative to $Q_{\Delta T}$. Notice, however, that by increasing the frequency the dissipative component represented by $\tilde{\vec{A}}$ in Eq. (\ref{w-a}) becomes more detrimental to the efficiency. There is, thus, a compromise between the two effects and an optimal frequency of operation.

\subsection{Driven quantum dot}
\label{example-qdot}

\begin{figure}[t]
\includegraphics[width=\columnwidth]{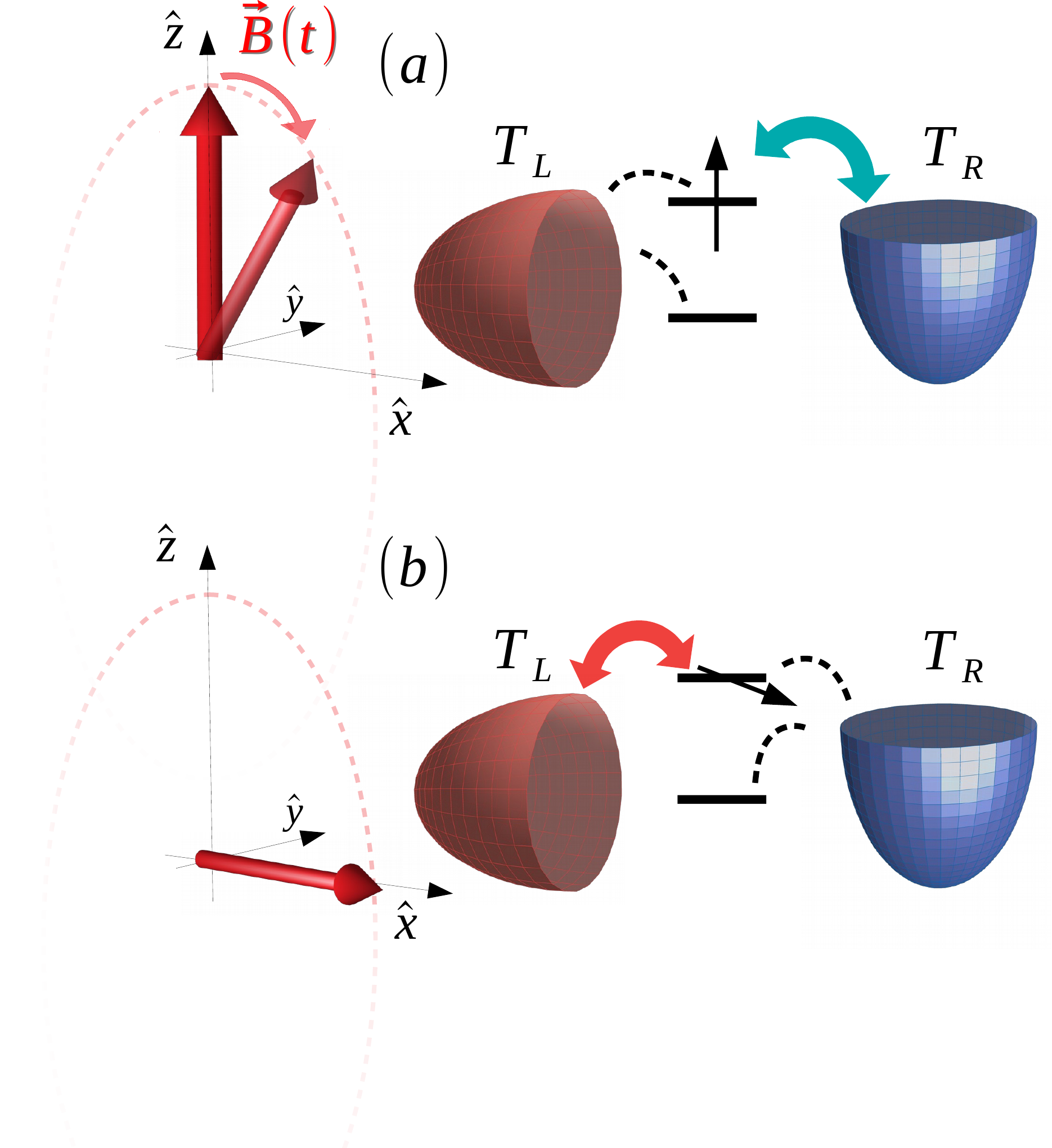}
\caption{Illustration of the quantum dot driven by a magnetic field and connected to electron reservoirs with different polarizations, represented by different orientations of the paraboloids.
The hybridization strength is modified according to the magnetic field's pointing direction. In (a) the electron hopping between the quantum dot
and the right ($z$-polarized) reservoir is favored, as is denoted by the thick arrow. In (b) the pointing direction of the magnetic field has changed to $x$ and now the quantum dot is stronger coupled to the left reservoir. }
\label{fig:qDotSchematic}
\end{figure}

In this case, the configuration consists of a central quantum dot driven by a time-dependent magnetic field and coupled to electron reservoirs with different polarizations. For the quantum dot the Hamiltonian ${\cal H}_S$ reads
\begin{equation}
\label{qdot}
{\cal H}_S(t)=  \; \Psi^{\dagger}_d  \left[  V_g \; \hat{\sigma}_0- \vec{B}(t) \cdot  \hat{\mbox{$\vec{\sigma}$}}  \right] \Psi_d,
\end{equation}
where $\Psi_d^{\dagger} = (d_{\uparrow}^{\dagger}, d_{\downarrow}^{\dagger})$ is a spinor related to the spin degrees of freedom of the electron in the quantum dot, while $d^{\dagger}_{\sigma}$ and $d_{\sigma}$ are respectively the creation and annihilation fermionic operators for these particles. The quantum dot contains two levels as a consequence of the Zeeman splitting introduced by the magnetic field. 
$\hat{\mbox{$\vec{\sigma}$}} = \left( \hat\sigma_x, \hat\sigma_y, \hat\sigma_z \right)$ is composed of the $2\times2$ Pauli matrices and $\hat{\sigma}_0$ is the identity, while
 $\vec{B}(t)= \left( B_x(t), B_y(t), B_z(t) \right)$ is the external time-periodic magnetic field
 and $V_g$ is a gate voltage, which rigidly
 shifts the energies of the two levels.

The reservoirs are represented by systems of non-interacting fermions.
The  electrons in the $\alpha$ reservoir are spin-polarized along the magnetization $\vec{m}_{\alpha}$. The Hamiltonian ${\cal H}_{\alpha}$ which describes the reservoir reads

\begin{equation}\label{dqres}
{\cal H}_{\alpha}=\sum_{k\alpha} \Psi_{k\alpha}^\dagger \left[ \varepsilon_{k \alpha} - \vec{ m}_{ \alpha} \cdot  \hat{\mbox{$\vec{\sigma}$}} \right]
\Psi_{k\alpha}, \;\;\; \alpha=L, R,
\end{equation}
where $\Psi_{k\alpha}^\dagger= \left( c_{k\alpha, \uparrow}^\dagger, c_{k\alpha, \downarrow}^\dagger \right)$ are spinors composed by the fermionic creation/annihilation operators $c_{k\alpha, \sigma}^\dagger$ and $c_{k\alpha,\sigma}$. We assume that both reservoirs have chemical potential $\mu_L=\mu_R=0$.

The coupling between the quantum dot and the reservoirs is represented by
\begin{equation}\label{qcontd}
{\cal H}_{c,\alpha}=\sum_{k\alpha,\sigma =\uparrow, \downarrow}V_{k\alpha, \sigma } \left( c^{\dagger}_{k\alpha, \sigma} d_{\sigma} + d_{\sigma}^{\dagger} c_{k\alpha, \sigma} \right).
\end{equation}
In order to solve the problem, it is convenient to change the basis of ${\cal H}_{\alpha}$
 to the one where the quantization axis for the spin coincides with the direction of $\vec{m}_{\alpha}$.
This is accomplished by the transformation
 $\left(c^{\dagger}_{k\alpha, \uparrow}, c^{\dagger}_{k\alpha, \downarrow}   \right) = \hat{U}^{\alpha} \left(c^{\dagger}_{k\alpha, +}, c^{\dagger}_{k\alpha, - }   \right)$.
In the new basis the Hamiltonians for the reservoirs and the couplings read
\begin{equation}\label{dqres1}
{\cal H}_{\alpha}=\sum_{k\alpha, s=\pm} c_{k\alpha,s}^\dagger \varepsilon_{k \alpha,s} c_{k\alpha,s}, \;\;\; \alpha=L, R,
\end{equation}
and 
\begin{equation}\label{qcontd1}
{\cal H}_{c,\alpha}=\sum_{k\alpha,s=\pm,\sigma =\uparrow, \downarrow} v_{k_{\alpha}s, \sigma}\left( c^{\dagger}_{k\alpha, s} d_{\sigma} + H. c. \right),
\end{equation}
with $ v_{k_{\alpha}s, \sigma}=U^{\alpha}_{s,\sigma} V_{k\alpha, \sigma }$.

 As discussed in Section \ref{ener-bal}, in order to have a non-vanishing pumping component we need to break spatial symmetry. We achieve this by considering different polarizations in the reservoirs. For concreteness, we consider the $L$ reservoir polarized along the positive $x$, and the $R$ one polarized along the positive $z$ direction. 
An illustration of the whole setup is sketched in Fig. \ref{fig:qDotSchematic}. 

This device bears resemblance to the driven q-bit discussed in Section \ref{example-qubit}. In fact, only the electrons with spins $z,\uparrow$ ($x, \uparrow$) 
can tunnel between the quantum dot and the $R$ ($L$) reservoir. Therefore,
when the magnetic field polarizes the quantum dot
along the positive $x$ direction, the tunneling of the electrons between the quantum dot and  the $L$ reservoir is optimal, while the tunnel between  the dot and the $R$ reservoir  is optimal when the
electron in the dot is polarized along the positive $z$ direction. The main difference between the present setup and the q-bit studied in Section \ref{example-qubit}  is the nature of the reservoirs, which is fermionic in the present case, while it is bosonic in the previous one. This difference is crucial from the technical point of view, because in the case of the quantum dot we will be able to solve the problem for arbitrary coupling between the driven system and the reservoirs. 
In addition, the quantum dot has a gate voltage, which moves its energy levels upwards or downwards in energy, thus tuning  different parts of the spectrum of the quantum dot into the relevant transport window --~$\sim k_B T$-- around the chemical potential of the reservoirs. This ingredient can be used to improve the performance, as we will discuss in Section \ref{q-dot-results}. Besides these differences, we expect the operation to be similar in both cases, at least within the regime where the coupling between the driven system and the reservoirs is very weak.

The heat-engine operational mode in the present case could be practically realized by implementing the time-dependent magnetic field by means of a rotating classical magnetic moment. 
The dynamics of the latter realizes the load of the heat engine. In such a case, a pumped heat $Q_{\rm tr,ac}$ flowing in the direction of the heat current induced by the thermal bias, will generate a torque and exert work on the magnetic moment, akin to the spin torque induced by an electrical bias \cite{torque1,torque2,torque3,torque4}. 

We will consider the same driving protocol as in the previous example, which is defined in Eq. (\ref{xb}), without focusing on the detailed mechanism generating the magnetic field. As in the previous example, we will show results for the heat pump and refrigerator modes.

\subsubsection{Green's function approach}

We can solve the problem exactly for arbitrary strength of the coupling between the quantum dot and the reservoirs by recourse to Green's functions. 
We will use the equilibrium finite-temperature formalism to evaluate the frozen susceptibilities  and  compute the response functions from Eq. (\ref{lambda}). This problem could be also exactly solved by recourse to the non-equilibrium Schwinger - Keldysh formalism in the Floquet representation and afterwards consider the expansion in small 
$\hbar \Omega$ and $\Delta T$ as in Refs. \cite{ludovico,lilimos} arriving at the same results as the ones we present here. We briefly summarize the  results below and show some details on the calculations  in Appendix \ref{aplam},
\begin{eqnarray}\label{lambdadot}
\Lambda^A_{3,\ell}(\vec{B})
&=& -\frac{1}{h}  \int d \varepsilon \frac{ d f(\varepsilon)}{d \varepsilon}  \varepsilon \mbox{Tr} \left[ \hat{\Gamma}_{R} \hat{\rho}(\varepsilon) \hat{\sigma}_\ell \hat{\rho} (\varepsilon) \right] , \;\ell=1,2\nonumber\\
\Lambda^S_{\ell,\ell'}(\vec{B})
&=& -\frac{1}{h} \int d \varepsilon \frac{ d f(\varepsilon)}{d \varepsilon} \mbox{Tr} \left[ \hat{\sigma}_{\ell} \hat{\rho} (\varepsilon) \hat{\sigma}_{\ell'} \hat{\rho}(\varepsilon) \right],\; \ell,\ell'=1,2 \nonumber\\
\Lambda^S_{3,3}(\vec{B})
&=& -\frac{1}{h} \int d \varepsilon \frac{ d f(\varepsilon)}{d \varepsilon}
\varepsilon^2 \mbox{Tr} \left[\hat{\Gamma}_{R}  \hat{G}_t (\varepsilon) \hat{\Gamma}_L \hat{G}_t^{\dagger}(\varepsilon) \right],
\end{eqnarray}
where $f(\varepsilon)=1/\left(e^{\varepsilon/(k_B T)} +1\right)$ is the Fermi-Dirac distribution function. We have also introduced the hybridization matrix
 $\hat{\Gamma}_{\alpha}$, with elements
\begin{equation}
(\hat{\Gamma}_\alpha)_{\sigma,\sigma'} = 2 \pi \sum_{k \alpha, s=\pm} U^{\alpha}_{\sigma,s} U^{\alpha}_{\sigma^{\prime},s} |V_{k\alpha}|^2 \delta (\varepsilon -\varepsilon_{k\alpha,s}).
\end{equation}
We consider  $L$ ($R$) reservoirs fully polarized with spins along the positive $x$ ($z$) directions
and a constant density of states. Thus,  $\Gamma_{\alpha} \simeq \sum_{k \alpha}  |V_{k\alpha}|^2 \delta (\varepsilon -\varepsilon_{k\alpha,+})$ and $\hat{\Gamma}_{\alpha} \simeq \Gamma_{\alpha} \hat{\tau}_{\alpha}$, with
\begin{equation}\label{gammatau}
\hat{\tau}_L \equiv \frac{1}{2} \left( \hat{\sigma}_x + \hat{\sigma}_0\right), \;\;\;\; \hat{\tau}_R \equiv \frac{1}{2} \left( \hat{\sigma}_z+ \hat{\sigma}_0 \right). 
\end{equation}
  The local density of states is described by the matrix 
  \begin{equation}
  \hat{\rho} (\varepsilon) =
 - 2 \mbox{Im}[\hat{G}_t(\vec{B},\varepsilon) ] = \hat{G}_t(\vec{B},\varepsilon)  \hat{\Gamma}\left[ \hat{G}_t(\vec{B},\varepsilon) \right]^{\dagger},
 \end{equation}
 which depends
 on the frozen Green's function
 \begin{equation}\label{gdot}
 \hat{G}_t(\vec{B},\varepsilon) = \left( \varepsilon - \vec{B}(t) \cdot  \hat{\mbox{$\vec{\sigma}$}} + i \hat{\Gamma}/2 \right)^{-1},
 \end{equation}
 with $\hat{\Gamma}= \hat{\Gamma}_L + \hat{\Gamma}_R$.

 In Eqs. (\ref{lambdadot}) we have highlighted the symmetric or antisymmetric nature of the components in each case. The fact that the components $\Lambda_{3,\ell}(\vec{B})$ are purely antisymmetric while 
 $\Lambda_{\ell,\ell^{\prime}}(\vec{B})$ are purely symmetric is a consequence of Onsager relations in combination with symmetry properties of the setup. These properties can be directly verified from the explicit calculations
 of Appendix \ref{aplam}. The last component $\Lambda^S_{3,3}(\vec{B})$ is proportional to the thermal conductance.
The symmetry properties  of $\Lambda_{\mu,\nu}(\vec{B})$ are the same as in the q-bit example of Section \ref{example-qubit}. Thus, the definitions of the vector potentials in the present case are the same as in Eq. (\ref{aas}).

\subsubsection{Results}\label{q-dot-results}

We carry out a similar analysis to the one  for the q-bit example given in Section \ref{example-qubit}. We consider the same two-parameter driving protocol as before, with  
$\vec{B}(t)=\left(B_x(t), 0, B_z(t)\right)$ given by Eq. (\ref{xb}).

\begin{figure}[t]
\centering
\includegraphics[width=0.9\columnwidth]{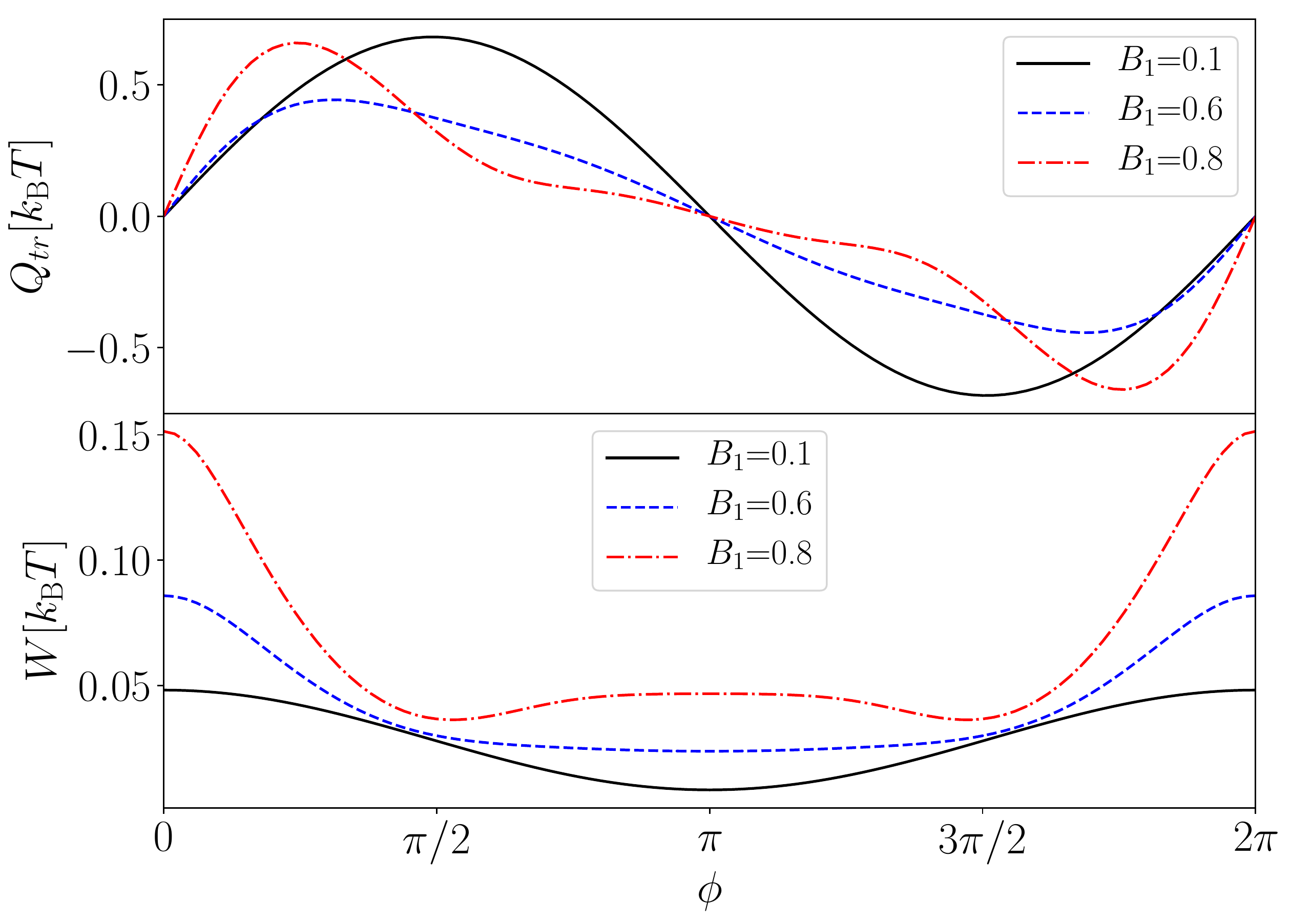}
\caption{ Pumped heat $Q_{\rm tr, ac} = Q_{\rm tr}$ (upper panel) and work done by the ac sources 
$W$ (lower panel) for $\Delta T=0$ as functions of the phase difference in the protocol defined by $B_x(t)= B_{x,0}+B_{x,1} \cos(\Omega t + \phi)$, $B_z(t)=B_{z,0}+B_{z,1} \cos (\Omega t)$ with  $B_{x,0} = B_{z,0}=0.4 k_B T$ and $B_{x,1}=B_{z,1}=B_1 k_B T $. $\Gamma_L=\Gamma_R=0.4 k_B T$ and $\hbar \Omega = k_B T/800$. The plot with $B_1=0.1$
 is multiplied by a factor $20$ in order to be shown in the same scale.}
\label{fig:pumping}
\end{figure}

As mentioned before, for the case of $V_g=0$ and weak coupling to the reservoirs, we expect a similar behavior to the case of the qubit. 
In Figure \ref{fig:pumping} we present the  pumped heat  $Q_{\rm tr, ac}(\phi)$ and the work developed by the ac sources $W$ for $\Delta T=0$,
as function of the driving phase difference $\phi$ between the two ac components of the magnetic field.  
As in the qubit case analyzed in
Section \ref{example-qubit},  for small amplitudes of the driving, $Q_{\rm tr, ac}$  is proportional to the area enclosed by the contour defined by the protocol. For this reason, the pumped heat 
behaves as $\propto \sin(\phi)$
and the generated work  as $\propto \cos(\phi)$ plus a constant. These functions are the same as in  the case of the driven qubit shown in Fig. \ref{fig:currlrs}. For larger values of the driving amplitude the pumped heat departs from this behavior.  However,   $Q_{\rm tr, ac}(\phi)$ vanishes for $\phi=0,\pi$ for any value of $B_{x,1}=B_{z,1}$. 

\begin{figure}[t]
\centering
\includegraphics[width=\columnwidth]{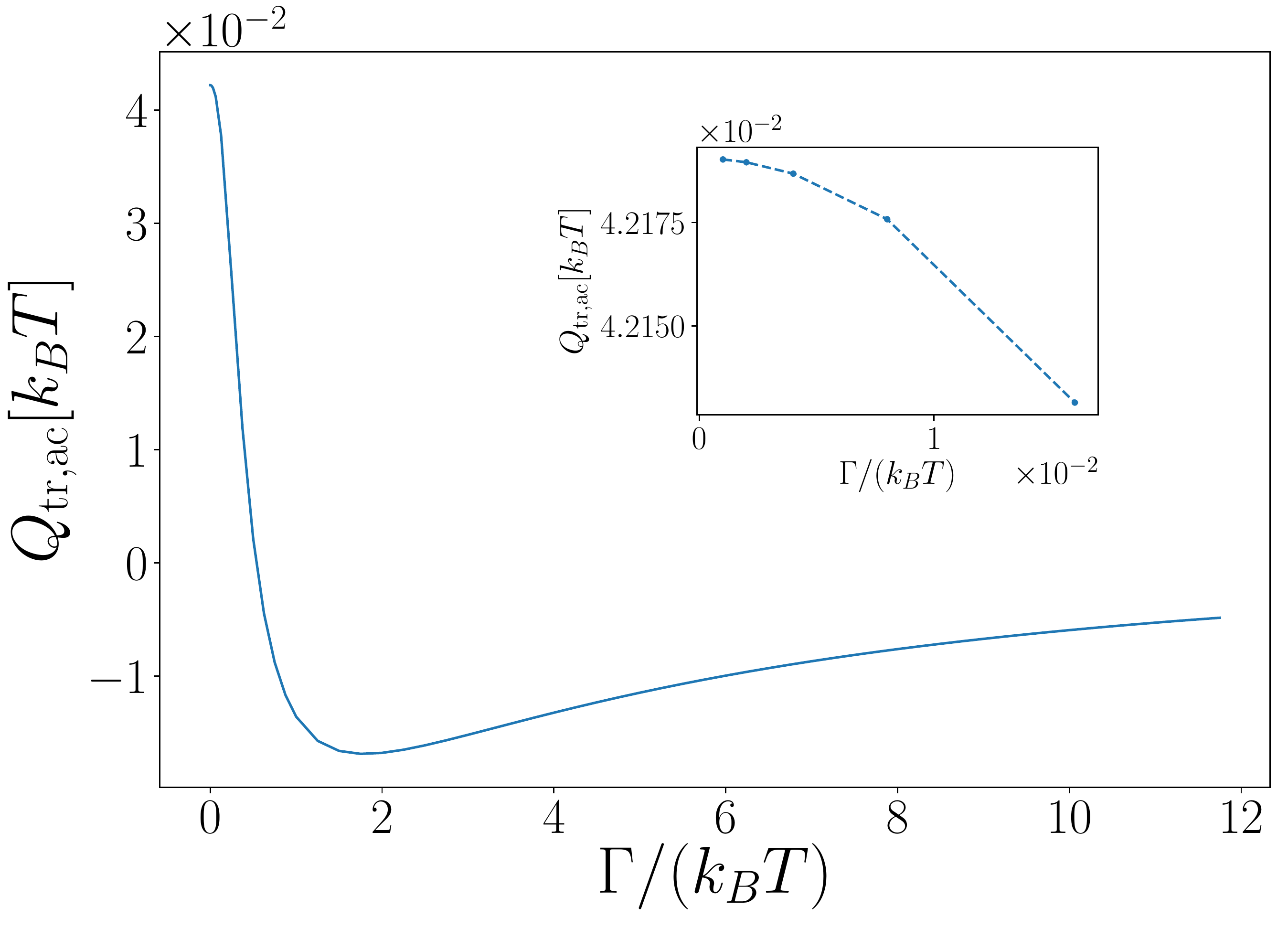}
\caption{Pumped heat $Q_{\rm tr, ac} = Q_{\rm tr}$ for the quantum dot with the same parameters as the q-bit operating with the protocol of Eq. (\ref{xb}) shown in 
 Fig. \ref{fig:currlrs} with $\phi=\pi/2$. 
}
\label{fig:weak}
\end{figure}

 In Fig. \ref{fig:weak} we further explore the comparison between the driven quantum dot and the driven q-bit. In particular, we
 show the behavior of the pumped heat as a function of the coupling to the reservoirs, assuming $\Gamma_L=\Gamma_R=\Gamma$ and the same parameters and driving protocol of
 Fig. \ref{fig:currlrs}. We can verify that as the latter parameter approaches the limit $\Gamma \rightarrow 0$, the value of the pumped heat of the  quantum dot  approaches the one of the qubit case shown in Fig. \ref{fig:currlrs}. There is some quantitative difference, which can be traced back to the fact that the type of couplings are not exactly the same (notice the matrix elements entering the couplings of the quantum dot are those of Eq. (\ref{gammatau}), while in the qubit we have considered $\hat{\sigma}_{x,z}$). 
 We  see that the strength of the coupling has a significant 
impact on the behavior of the pumped heat. For the present parameters, we observe an inversion in the direction of the pumped heat as the coupling increases and overcomes 
$\Gamma \sim |\vec{B}|$, at which the width of the levels of the quantum dot becomes comparable to the energy difference between them. 

 \begin{figure*}[t]
\centering
\includegraphics[width=\textwidth]{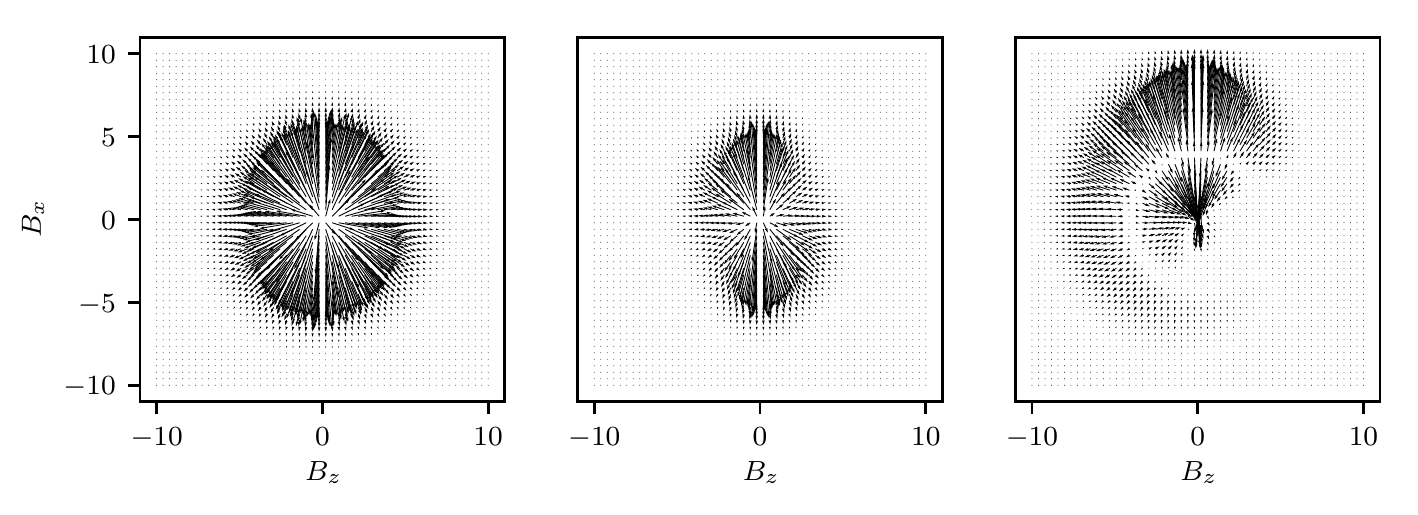}
\caption{Vector fields $\vec{A}^A_3(\vec{B})$  for a driving protocol with $\vec{B}(t)=\left(B_x(t),0,B_z(t)\right)$, corresponding to a quantum dot coupled to reservoirs with different polarizations. $R$ ($L$)  reservoirs is polarized along positive $z$ ($x$) direction. Left panel corresponds to 
$\Gamma_L=\Gamma_R$ and $V_g=0$, middle  correspond to $\Gamma_L=0.1 \Gamma_R$ and $V_g=0$, while 
right panel corresponds to $\Gamma_L=0.1 \Gamma_R$ and $V_g= 2 \Gamma_R$. 
 The temperature of the reservoirs is $k_B T=0.5 \Gamma_R$.}
\label{fig:flor}
\end{figure*}

\begin{figure}
\centering
\includegraphics[width=\columnwidth]{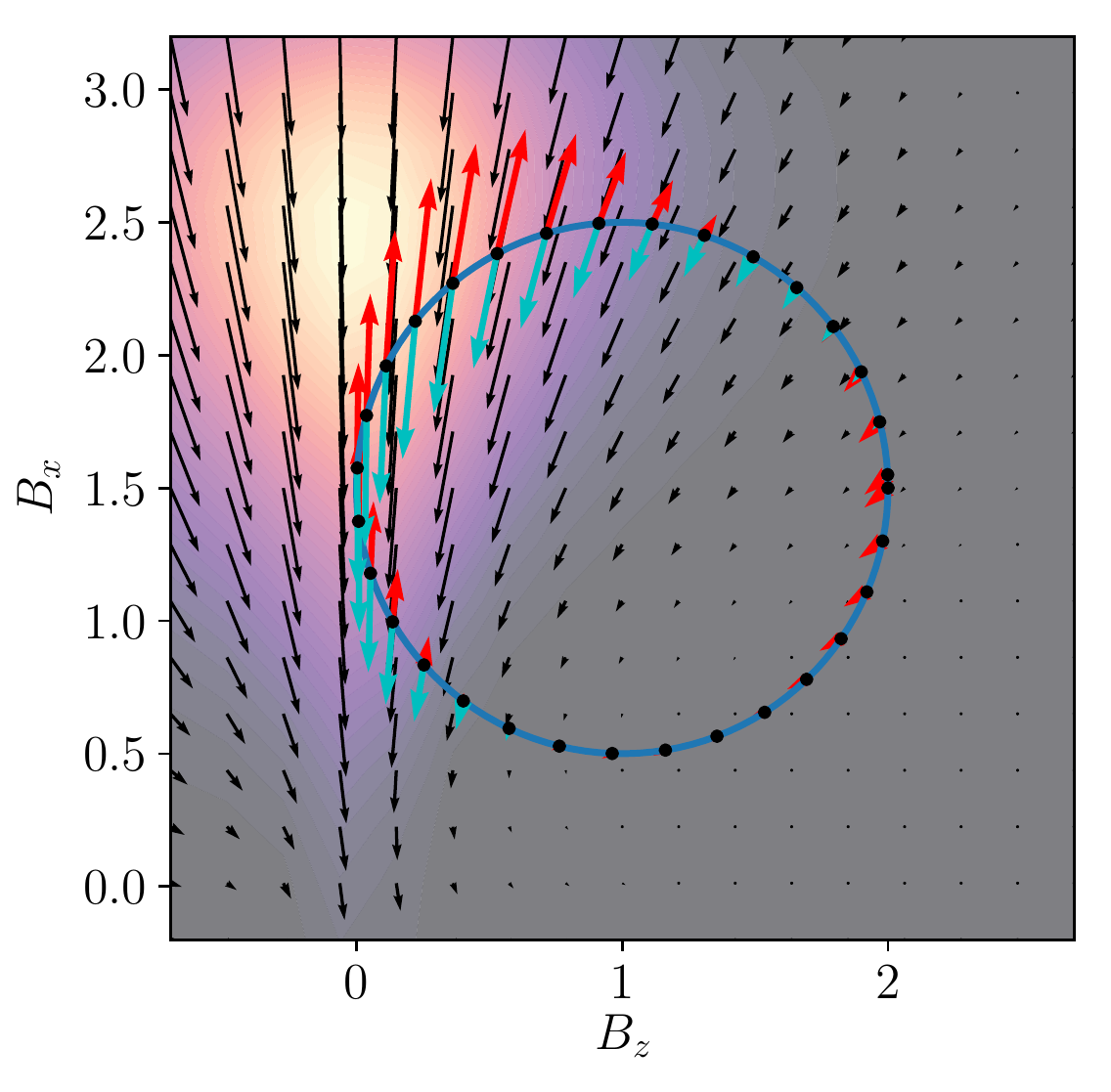}
\caption{Vector fields  $\vec{A}^A_3(\vec{B})$ (cyan) and $\tilde{\vec{A}}^S(\vec{B})$ (red) over a closed path (solid blue curve) for the configuration shown in the lower panel of Fig. \ref{fig:flor} 
($\Gamma_R=0.1 \Gamma_L$). The
driving protocol defining the path is  $B_x(t)= B_{x,0}+B_{x,1} \cos(\Omega t + \phi)$, $B_z(t)=B_{z,0}+B_{z,1} \cos (\Omega t)$ with $B_{x,0}=1.5 \Gamma_R,\; B_{z,0}= \Gamma_R$, 
$B_{x,1}=B_{z,1}= \Gamma_R$, 
$\phi= \pi/2$. The black arrows represent $\vec{A}_3(\vec{B})$ outside the defined protocol.}
\label{fig:protocol}
\end{figure}

\begin{figure}
\includegraphics[width=\columnwidth]{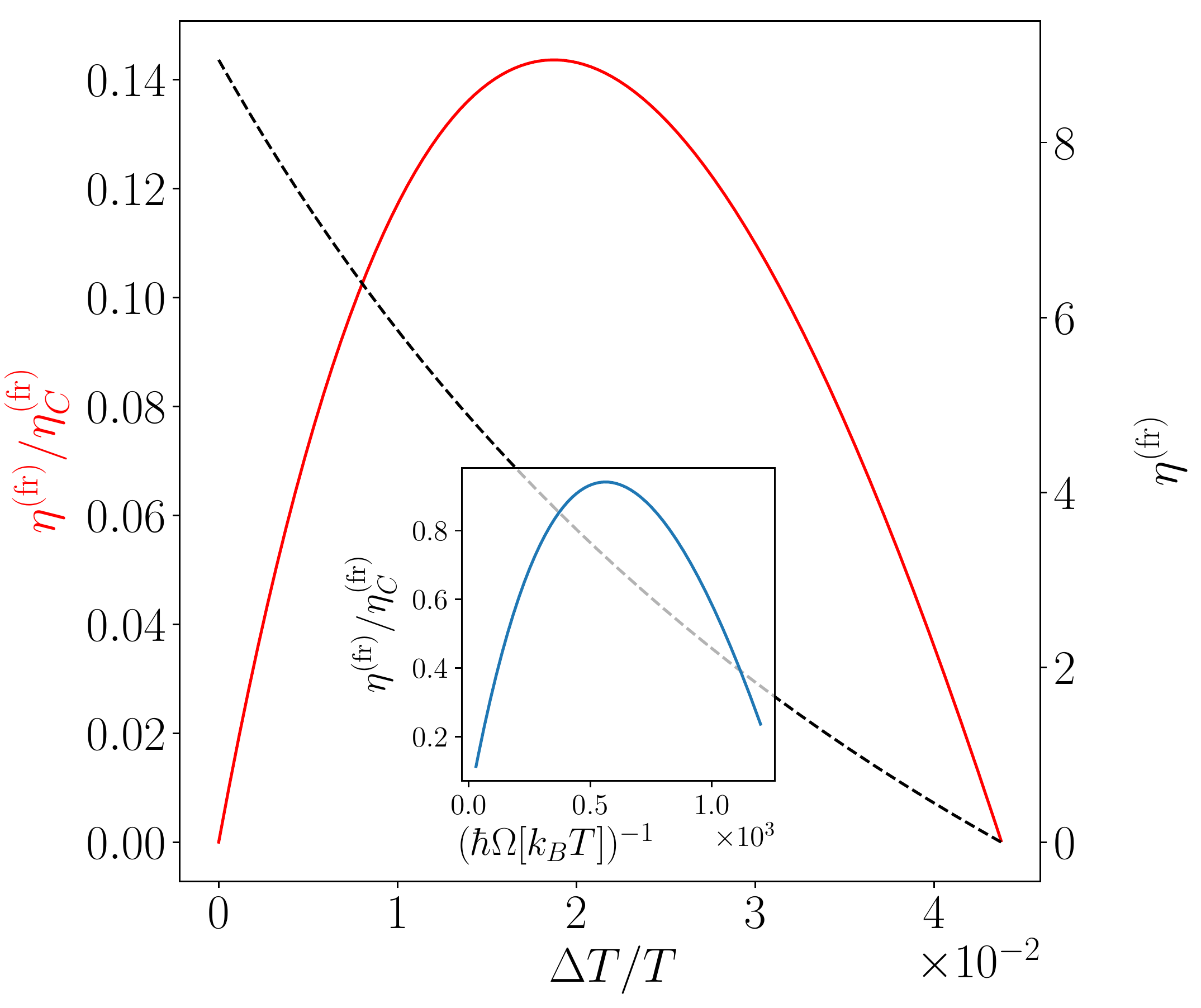}
\caption{Coefficient of performance for refrigeration (absolute in dashed black and normalized to the Carnot value in red) versus $\Delta T$ for the protocol of  Fig. \ref{fig:protocol}
for $\hbar \Omega= \Gamma_R/200$ Inset: Normalized coefficient of performance for refrigeration as a function of $\hbar \Omega$ for $\Delta T = T/150 $.}
\label{fig:eff_dT}
\end{figure}

We now focus on the properties in the operation  of the quantum-dot machine that are different from the weakly coupled driven q-bit.
To this end, we further analyze  the structure of 
 the vector potentials $\vec{A}_{\mu}^{S/A}(\vec{B})$ and $\tilde\vec{A}^{S/A}(\vec{B})$ in Eq. (\ref{aas}) with the tensor 
$\Lambda_{\mu,\nu}(\vec{B})$ of Eq. (\ref{lambdadot}). The vector map for $\vec{A}^A_3(\vec{B})$ in the parameter space for a given temperature $T$ is shown in Fig. \ref{fig:flor}.
This representation is useful to visualize the symmetries of the setup and to select the driving protocol that maximizes the contour integral
$\oint \vec{A}^A_3(\vec{B}) \cdot d\vec{B}$.
 In the left panel the quantum dot is contacted with the same strength to both reservoirs ($\Gamma_L=\Gamma_R$),  $L$ being polarized along positive $x$ and $R$ along positive $z$ direction, as
 indicated in the sketch of Fig. \ref{fig:qDotSchematic}.  In the middle panel, the contact  is stronger  to $L$ than to 
$R$ ($\Gamma_R=0.1 \Gamma_L$). Consequently, we can visualize a higher intensity of the field $\vec{A}_3^A$ along the $B_x$ than along the $B_z$ direction. Both left and middle plots have $V_g=0$, in which case 
the Hamiltonian of  Eq. (\ref{qdot}) is symmetric under the simultaneous transformations
$\Psi^{\dagger}_d \rightarrow \Psi_d$ and $\vec{B} \rightarrow -\vec{B}$. The first one is a particle-hole transformation, under which  the heat current changes the sign.
Consequently, 
the field maps of Fig. \ref{fig:flor} present the symmetry $\vec{A}^A_3(\vec{B})=-\vec{A}_3^A(-\vec{B})$. In the right panel, we can visualize that the  breaking of the particle-hole symmetry
by a gate voltage introduces a strong asymmetry in the vector field.

 With the picture of Fig. \ref{fig:flor} in mind, we can readily design a closed trajectory that optimizes pumping. The latter
corresponds to a path that goes parallel to the vector field  within the region where its  intensity is high, and closes antiparallel to the vector field in a 
very low-intensity region. An example of such a trajectory is shown in Fig. \ref{fig:protocol}.
The corresponding vectors $\tilde\vec{A}^S(\vec{B})$ along the trajectory are also shown in cyan. Trajectories leading to high efficiencies of the machine would have as small dissipation as possible, in addition to high values of heat pumping. While the optimization of the pumping can be easily achieved by recourse to the vector field representation $\vec{A}^A_3(\vec{B})$, it is not
easy to optimize a trajectory to decrease the integral over $\tilde\vec{A}^S(\vec{B})$. However, we know that this quantity can be reduced by decreasing the pumping frequency $\Omega$. 

In Fig. \ref{fig:eff_dT} we illustrate the behavior of the COP of the driven quantum dot operating as a refrigerator. Overall, this quantity follows a similar behavior as a function of 
$\Delta T/T$ and $\Omega$ as the one of the qubit (see Fig. \ref{fig:heatpmpcopqbtdt2}). Therefore, most of the comments and remarks presented in the analysis of Fig.~\ref{fig:heatpmpcopqbtdt2} apply also here.
 However, it is several orders of magnitude higher in the present case, achieving values as large as
$14 \; \%\; \eta_C^{\rm fr}$. The key for this improvement is the selection of an appropriate pumping protocol, taking advantage of the extra features introduced by the existence of the gate voltage
 $V_g$ in the present problem.

\begin{figure}[t]
\centering
\includegraphics[width=\columnwidth]{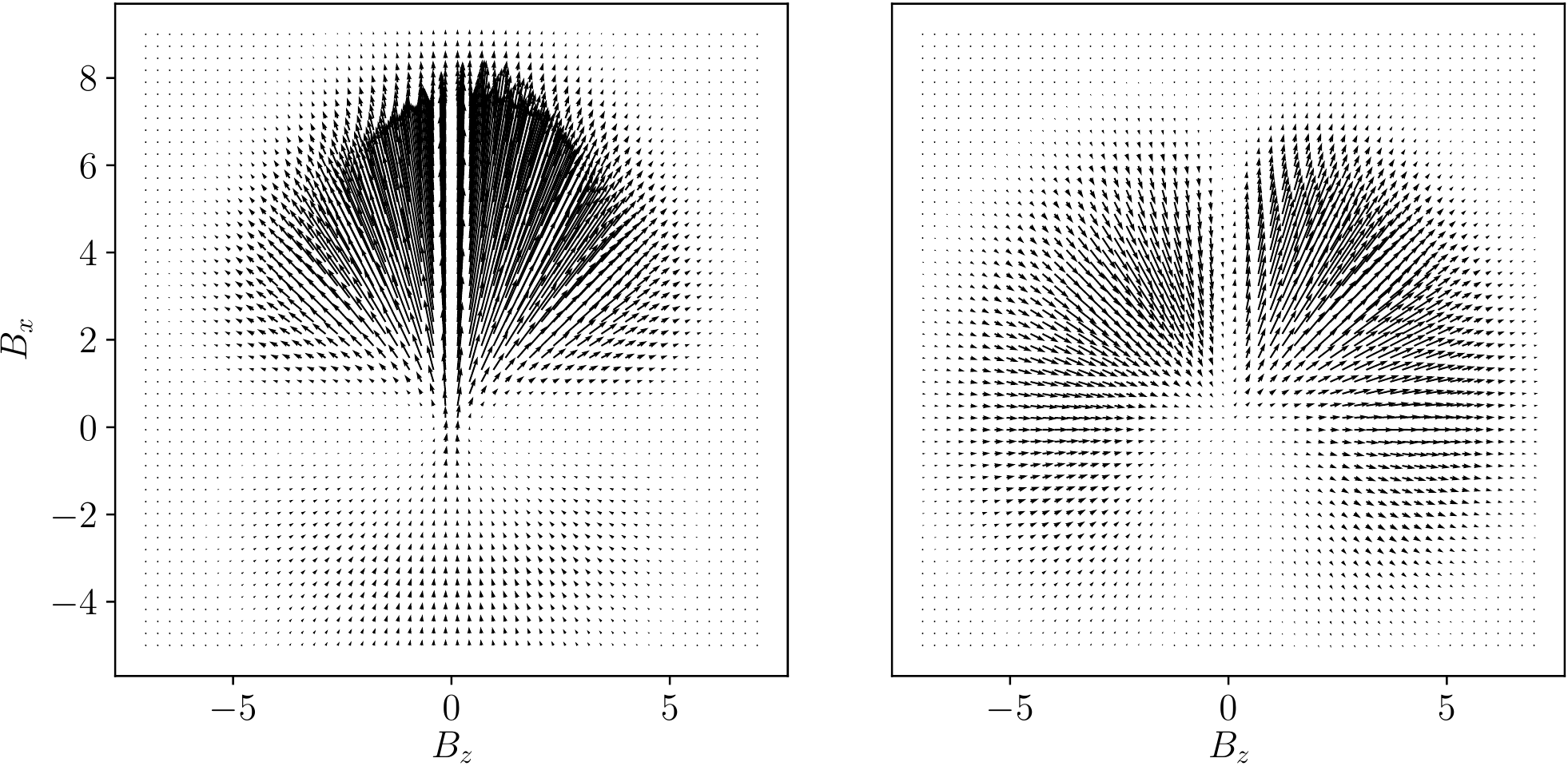}
\caption{Vector fields $\vec{A}^S_1(\vec{B})$ (left) and $\vec{A}^S_2(\vec{B})$ (right) following Eqs.~(\ref{aas})
and (\ref{lambdadot}), for the parameters of the right panel of Fig. \ref{fig:flor}. 
}
\label{fig:mag}
\end{figure}

\begin{figure}[t]
\centering
\includegraphics[width=\columnwidth]{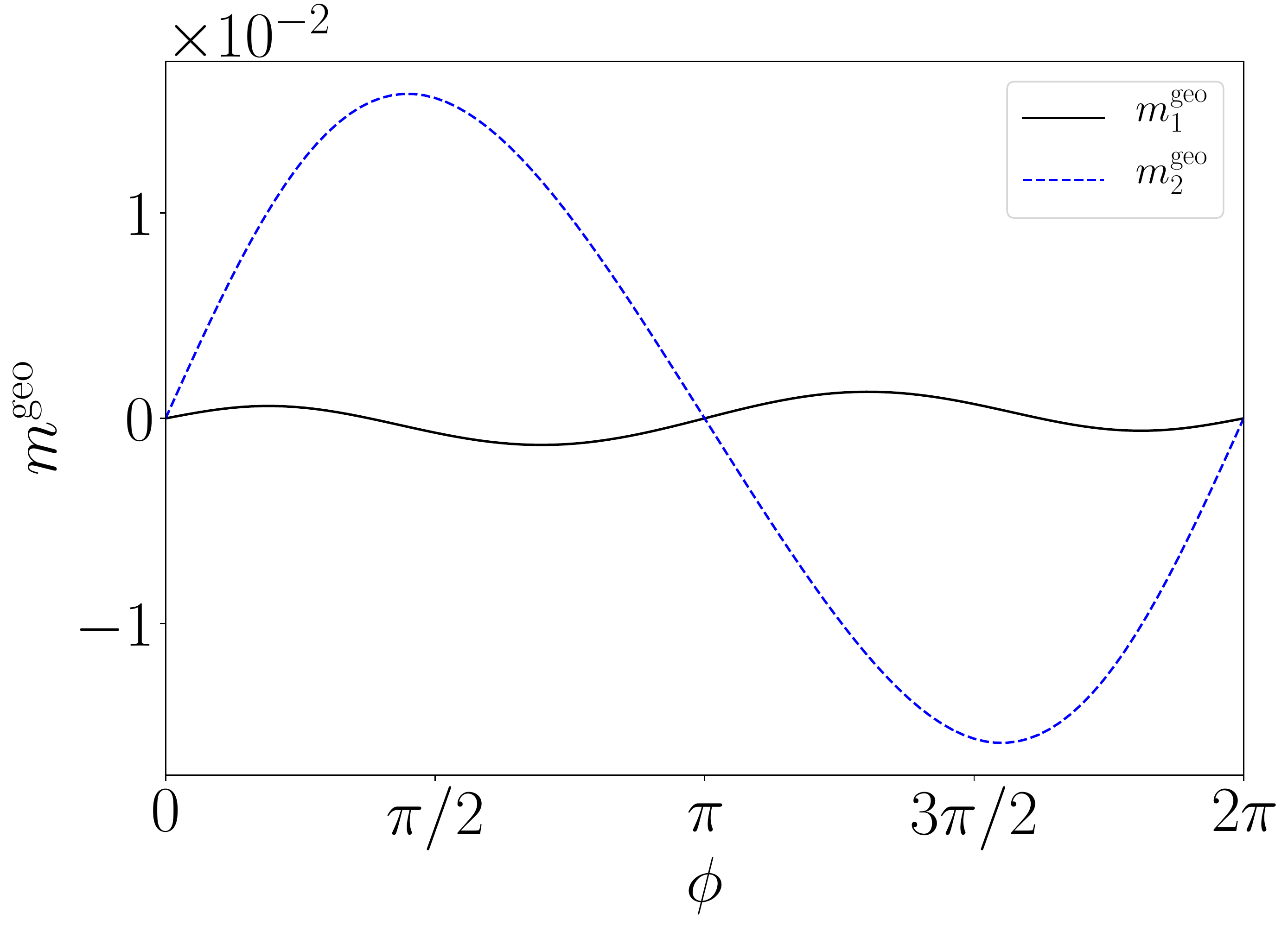}
\caption{Components of the geometric magnetization $m_{1,2}^{\rm geo}$, defined in Eq. (\ref{n12}) as functions of the phase-lag $\phi$ corresponding to paths of the form $B_x(t)= B_{x,0}+B_{x,1} \cos(\Omega t + \phi)$, $B_z(t)=B_{z,0}+B_{z,1} \cos (\Omega t)$ with $ B_{x,0}=1.5 \Gamma_R,\; B_{z,0}= \Gamma_R$, $B_{x,1}=B_{z,1}= \Gamma_R$, on the vector fields of Fig. \ref{fig:mag}.
}
\label{fig:magphi}
\end{figure}

We close this section by analyzing the geometric component of the first-order adiabatic reaction force defined in Eq. (\ref{avforceline}). In the present problem, the latter coincides with the
magnetic moment of the quantum dot.  For $\Delta T=0$, the magnetic moment of the quantum dot is given by
\begin{eqnarray}\label{n12}
m_{\ell} &=&=\frac{\Omega}{2\pi} \int_0^{2 \pi/\Omega} dt \langle \Psi^{\dagger}_d \hat{\sigma}_{\ell} \Psi_d \rangle (t) = m^{\rm BO}_{\ell}+m^{\rm geo}_{\ell},
\nonumber \\
m^{\rm geo}_{\ell}&=& \frac{\Omega}{2\pi} \oint \vec{A}^S_{\ell}(\vec{B}) \cdot d \vec{B},
\end{eqnarray}
with $\hat{\sigma}_{x,y}$ for $\ell \equiv 1,2$, respectively.  Here, $m^{\rm BO}_{\ell}$ is the average over one period of the the instantaneous magnetization corresponding to the equilibrium frozen Hamiltonian,
while $m^{\rm geo}_{\ell}$ is the geometric component, corresponding to the first-order adiabatic reaction force of Eq. (\ref{avforceline}). The vectors $\vec{A}^S_{\ell}(\vec{B})$ are calculated from
Eq. (\ref{lambdadot}) as defined in Eq. (\ref{aas}).  Interestingly,
the symmetric component of the thermal geometric tensor, which defines the dissipation, is directly related in the present problem to a local physical quantity, which is the quantum dot geometric magnetization \cite{pablo}.
The latter is experimentally accessible. In fact, notice that the component $m^{\rm BO}_{\ell}$ does not explicitly depend on the driving frequency, while the second term has an explicit linear dependence on $\Omega$.
Therefore, in a concrete experimental measurement of the quantum dot magnetization, both components should be distinguishable from one another. 

The associated vector fields $\vec{A}^S_{\ell}(\vec{B})$ are shown in Fig.~\ref{fig:mag} for configurations with stronger coupling to the  $L$ ($x$-polarized) reservoir than to the $R$ ($z$-polarized) one and a finite gate voltage $V_g$, with the same values of the parameters as in the right panel of Fig.~\ref{fig:flor}. In this representation, we can visualize higher intensity of the fields along $B_x, B_z >0$ relative to $B_x, B_z <0$, as a consequence of the polarization of the reservoirs along the positive $x$ and $z$-axis. 
The amplitudes of $\vec{A}^S_1(\vec{B})$, shown in the left panel, are larger than those of $\vec{A}^S_2(\vec{B})$, shown in the right panel, due to the larger coupling to the reservoir polarized along $x$. The result of calculating the integrals over closed
trajectories  with different phase lags $\phi$ between the components $B_x$ and $B_z$ is shown in Fig. \ref{fig:magphi}. As in the case of the pumped heat, both components of the magnetization vanish at $\phi=0,\pi$. 

\section{Summary and conclusions}
\label{conclusions}
We have presented a general description of the geometrical properties of quantum thermal machines under the effect of adiabatic periodic driving and a small thermal bias due to the contact to reservoirs at different temperatures.
The cyclic time-dependence is introduced via classical variables, varying slowly in time, that enter the quantum Hamiltonian of the 
system.
We show that the operation of the thermal machine, consisting of a few-level quantum system, is fully characterized by the thermal tensor $\Lambda_{\mu,\nu}$ defined in Section~\ref{geo-section}.

The formal derivation of this tensor is obtained by means of the adiabatic linear response theory complemented by Luttinger's representation of the thermal bias.
The symmetric component of $\Lambda_{\mu,\nu}$ characterizes the total rate of entropy production, thus controlling the dissipation of all the sources involved in the operation of the machine.
When the system is driven by two or more periodically-varying parameters, it is possible to obtain pumping of heat between reservoirs, even in the absence of a temperature bias.
The heat pumped, the work performed on the system, and the dissipated power can be described by means of vector fields defined through the thermal tensor.
In particular, the pumped heat by the driving and the work performed can be expressed in a purely geometric form as line integrals of those vector fields over the closed paths which represent the driving cycles in the parameter space.
In the presence of a thermal bias, these two quantities allow the characterization of a thermal machine which realizes heat-work conversion.

We have illustrated these ideas using two paradigmatic quantum systems coupled to two thermal reservoirs.
The first example consists of a qubit, whose energy levels and inter-level tunneling depend harmonically on time, attached to two bosonic reservoirs kept at different temperatures.
The second example is a quantum dot coupled to electronic reservoirs and driven by a harmonically time-dependent and rotating magnetic field.
The two examples are solved with different techniques, while two driving parameters are assumed. In the case of the qubit we rely on the master equation approach, valid for weak coupling to the reservoirs, while in the case of the quantum dot we solve the problem exactly for arbitrary coupling by recourse to  linear response and Green's function formalisms.
The two problems are very similar qualitatively and quantitatively when the driven system is weakly coupled to the reservoirs.
In the two cases, we have calculated the vector fields responsible for the geometric characterization of the systems as thermal machines.
We have computed the heat pumped and the work as functions of: i) phase lag between the two driving parameters, ii) the reference temperature, and iii) the coupling between system and reservoir (for the second example).
The efficiency of the thermal machines has been analyzed in terms of the temperature difference between reservoirs, the average temperature, and the frequency of the driving parameters in both cases. 
Finally, in the second example, we have shown how the representation of the pumped heat by means of vector fields can be used to identify the cycles that maximize it, thus improving the performance of a thermal machine.

\section{Acknowledgements}
We acknowledge support from PIP-2015-CONICET,  PICT-2014, PICT-2018, Argentina (PTA and LA), Simons-ICTP-Trieste associateship  and the Alexander von Humboldt Foundation, Germany (LA),
Deutsche Forschungsgemeinschaft through CRC 183 (FvO),
ICTP Federation agreement (PTA), grant CNR-CONICET (BB, LA and FT), the Oxford Martin Programme (RF). We thank the Dahlem Center for Complex Quantum Systems, Berlin and the International Center for Theoretical Physics, Trieste, for hospitality.



\appendix
\section{Luttinger theory of thermal transport}
\label{Lutt-theory}
The idea of expressing the thermal difference in a Hamiltonian language was originally introduced by Luttinger \cite{luttinger}. Here, we   follow the revised 
version of Luttinger's theory presented by Tatara in Ref.~\onlinecite{tatara}, which we briefly review and adapt in order to deal with a Hamiltonian containing a 
tunneling contact between the central system and the reservoirs at which the thermal difference is applied. Luttinger's theory is formulated in the continuum 
starting from a Hamiltonian ${\cal H}_{\rm E}(t)= \int \mathrm{d}{\bf r} h({\bf r}) \psi({\bf r}, t) $, where $ \psi({\bf r}, t) $ is a "gravitational" potential. Gradients of the latter  
induce energy flows ${\bf j}^E$ akin to the electrical currents induced by gradients of the electric potential. Such energy flows obey a continuity equation 
$\dot{ h}({\bf r}) =- \partial_{\bf r} \cdot {\bf j}^E({\bf r})$ as a consequence of energy conservation, which motivates the definition 
\begin{equation}\label{lut}
	{\cal H}_{\rm Lutt}(t) = \int_{-\infty}^t \mathrm{d}t^{\prime} \int \mathrm{d}{\bf r} \; {\bf j}^E(t^{\prime} ) \cdot \partial_{\bf r} \psi({\bf r}, t),
\end{equation}
with $ \partial_{\bf r} \psi({\bf r}, t)= \partial_{\bf r} T/T$. 
Such formulation is consistent with the rate of change of the entropy production,
\begin{equation}
	\dot{S}= - \int \mathrm{d}{\bf r}  \frac{1}{T} \partial_{\bf r}\cdot  \langle {\bf j}^E(t) \rangle=  - \int \mathrm{d}{\bf r}  \langle{\bf j}^E(t)\rangle \cdot   \frac{\partial_{\bf r} T}{T^2},
\end{equation}
through the relation $\langle {\cal H}_{\rm Lutt}(t) \rangle=  T S$.

Ref.~\onlinecite{tatara} considers the alternative Hamiltonian
\begin{equation}\label{at}
{\cal H}_{ A_T}(t) = - \int d{\bf r} \; {\bf j}^E(t^{\prime} ) \cdot \vec{A}_T({\bf r},t). 
\end{equation}
The Hamiltonians of Eqs. (\ref{lut}) and (\ref{at}) coincide in the long-time average. In fact, $\int_{-\infty}^{+\infty} \mathrm{d}t \; {\cal H}_{\rm Lutt}(t)= 
\int_{-\infty}^{+\infty} \mathrm{d}t \; {\cal H}_{A_T}(t)$ with 
\begin{equation}\label{adel}
\partial_t \vec{A}_T({\bf r},t)= \partial_{\bf r} \psi({\bf r}, t)= \partial_{\bf r} T/T.
\end{equation}
 In this way, $\vec{A}_T({\bf r},t)$ and $\psi({\bf r}, t)$
behave, respectively, in a similar way as the vector and scalar potentials of electromagnetism.

\section{Identities satisfied by the adiabatic susceptibilities}\label{relat}
In order to prove the identities of Eq. (\ref{ident}), satisfied by the adiabatic susceptibilities for the thermal driving corresponding to the {\em frozen} Hamiltonian ${\cal H}_t$,  we proceed by writing the
following equation satisfied by the current operators,
\begin{equation}
 {\cal J}^E_L(t) + {\cal J}^E_R(t) = \dot{\cal H}_S(t),
\end{equation}
where $\dot{\cal H}_S$ encloses all the terms of ${\cal H}_t$ corresponding to the central system and contacts between system and reservoirs.
All the operators are expressed in Heisenberg representation with respect to ${\cal H}_t$ 
\begin{equation}\label{eqlr}
\sum_{\alpha, \beta=L,R} \chi_t^{\rm ad} \left[ {\cal J}^E_{\alpha} , {\cal J}^E_{\beta}\right] = \chi_t^{\rm ad} \left[  \dot{\cal H}_S,  \dot{\cal H}_S\right]=0.
\end{equation}
In order to  prove that the right-hand side (rhs) of this equation is zero we start  from the definition of the adiabatic susceptibility,
\begin{equation}\label{chiadlim}
 \chi_t^{\rm ad} \left[  \dot{\cal H}_S,  \dot{\cal H}_S\right]=-i \lim_{\omega \rightarrow 0} 
 \partial_{\omega} \chi_{\dot{S},\dot{S}}(\omega)  = \lim_{\omega \rightarrow 0} \frac{\mbox{Im}[\chi_{\dot{S},\dot{S}}(\omega)]}{\omega},
\end{equation}
being  $\chi_{\dot{S},\dot{S}}(\omega)$ the Fourier transform of 
 the susceptibility $\chi_{\dot{S},\dot{S}}(t-t^{\prime})=-i\theta(t-t^{\prime}) \langle \left[  \dot{\cal H}_S(t),  \dot{\cal H}_S(t^{\prime})\right]\rangle_t$. Since all the mean
 values correspond to the equilibrium frozen Hamiltonian ${\cal H}_t$, we have $\chi_{\dot{S},\dot{S}}(t-t^{\prime})= \partial_t \partial_{t^{\prime}} \chi_{S,S}(t-t^{\prime})$, being 
  $\chi_{S,S}(t-t^{\prime})=-i\theta(t-t^{\prime}) \langle \left[  {\cal H}_S(t),  {\cal H}_S(t^{\prime})\right]\rangle_t$. Hence, 
  \begin{equation}\label{chids}
  \chi_{\dot{S},\dot{S}}(\omega) = - \omega^2 \chi_{S,S}(\omega).
  \end{equation}
  For a system with a bounded spectrum,  $\chi_t^{\rm ad} \left[  \dot{\cal H}_S,  \dot{\cal H}_S\right] =0$ when the limit $\omega \rightarrow 0$
  is evaluated in Eq. (\ref{chiadlim}). In fact, introducing  the Lehmann representation in $\chi_{S,S}(\omega)$
  and using (\ref{chids}) and  (\ref{chiadlim}) we get
 \begin{eqnarray} \label{chissd}
& & \chi_t^{\rm ad} \left[  \dot{\cal H}_S,  \dot{\cal H}_S\right] =\pi \lim_{\omega \rightarrow 0}  \omega^2 \sum_{n,m }  p_m |\langle m|H_S | n\rangle|^2\nonumber \\
 & &\;\;\;\;\;\;\; \times \left[\delta(\omega-(\varepsilon_m-\varepsilon_n))- \delta(\omega-(\varepsilon_n-\varepsilon_m))\right] 
 \end{eqnarray}
with ${\cal H}_t |m\rangle = \varepsilon_m |m\rangle$. In the latter equation $ |\langle m|H_S | n\rangle|^2$ is finite for a system with a bounded spectrum, while 
$\sum_{n,m }\left[\delta(\omega-(\varepsilon_m-\varepsilon_n))- \delta(\omega-(\varepsilon_n-\varepsilon_m))\right] $ is the density of states for the excitations of the full system.
Typically, the latter function is gapped or has a power-law behavior $\sim |\omega|^{\gamma}$ with $\gamma >0$, which proves the rhs of Eq.~(\ref{eqlr}).

Using Eq.~(\ref{eqlr}), we get the  identities of Eq.~(\ref{ident}). A similar argument can be elaborated for the identities
related to the response functions combining energy currents and ac-driving forces. In that case, we can prove
\begin{eqnarray}
& & \sum_{\alpha=L,R} \chi_t^{\rm ad} \left[ {\cal J}^E_{\alpha} , {\cal F}_l\right] = \chi_t^{\rm ad} \left[  \dot{\cal H}_S,  {\cal F}_l \right]=0,\nonumber \\
& & \sum_{\alpha=L,R} \chi_t^{\rm ad} \left[ {\cal F}_l, {\cal J}^E_{\alpha} \right] = \chi_t^{\rm ad} \left[ {\cal F}_l,  \dot{\cal H}_S \right]=0,
\end{eqnarray}
following similar reasoning as with Eq. (\ref{eqlr}).

Summarizing, the adiabatic response functions  in which the energy current enters are
  \begin{eqnarray}\label{ident}
  	\chi_t^{\rm ad} \left[ {\cal J}^E_{\alpha} , {\cal J}^E_{\alpha}\right] &=& \chi_t^{\rm ad}\left[ {\cal J}^E_{\bar{\alpha}} , {\cal J}^E_{\bar{\alpha}}\right] 
	 \nonumber \\
	\chi_t^{\rm ad} \left[ {\cal J}^E_{\alpha} , {\cal J}^E_{\alpha}\right] &=& - \chi_t^{\rm ad}\left[ {\cal J}^E_{\alpha} , {\cal J}^E_{\bar{\alpha}}\right],  \\
\chi_t^{\rm ad} \left[  {\cal F}_{l}  , {\cal J}^E_{\alpha} \right] &=& - \chi_t^{\rm ad} \left[  {\cal F}_{l}  , {\cal J}^E_{\bar{\alpha}}\right],  \nonumber\\
\chi_t^{\rm ad} \left[ {\cal J}^E_{\alpha}, {\cal F}_{l}   \right] &=& - \chi_t^{\rm ad} \left[ {\cal J}^E_{\bar{\alpha}}, {\cal F}_{l}   \right],\label{fj}
\end{eqnarray}
up to some function that vanishes when averaging over one period. In the above equations $\bar{\alpha}$ denotes the reservoir opposite  to $\alpha$.

\section{Entropy production rate}
\label{entropy-rate}
In what follows, we present a microscopic derivation of the expression for the entropy production rate associated to the combined effect of the time-dependent
and thermal driving in the adiabatic regime.
\subsubsection{ac driving}
We start by analyzing the effect of the time-dependent driving. 
To this end,  we can proceed along the lines of Refs. \cite{deff,ludo} and start from the definition of von Neumann entropy
\begin{equation}\label{vn}
S(t)= -k_B \mbox{Tr}\left[ \rho(t) \mbox{ln} \rho(t) \right].
\end{equation}
We also introduce the following auxiliary function,
\begin{equation}
S[{\cal H}_t]=  k_B \mbox{Tr}\left[ \rho(t)  \left( \beta{\cal H}_t +\mbox{ln}Z_t \right) \right],
\end{equation}
 with 
 $Z_t= \mbox{Tr}\left[e^{-\beta {\cal H}_t}\right]$. Under a 
 small change  in the parameter space, ${\bf X}(t) \rightarrow {\bf X}(t+\delta t)$, the Hamiltonian evolves to 
 \begin{equation}
  {\cal H}_{t+\delta t} = {\cal H}_{t}+
 \frac{\partial {\cal H}(t)}{\partial {\bf X} } \cdot \dot{\bf X}(t) \delta t =
{\cal H}_{t}- \mathbfcal{F}\cdot \dot{\bf X}(t) \delta t.
 \end{equation}
 Consequently,
 \begin{equation}
S[{\cal H}_{t+\delta t}]=  k_B \mbox{Tr}\left[ \rho(t+\delta t)  \left( \beta{\cal H}_{t+\delta t} +\mbox{ln}Z_{t+\delta t} \right) \right].
\end{equation}
 The  change  in the latter function is
 $\delta S[{\cal H}]=S[ {\cal H}_{t+\delta t}]-S[ {\cal H}_{t}]$, which keeping terms up to first order in $\delta t$ explicitly reads
\begin{eqnarray}
& & \delta S[{\cal H}]= k_B \beta \left\{  \mbox{Tr}\left[ \rho(t+\delta t) {\cal H}_t \right] -    \mbox{Tr}\left[ \rho(t) {\cal H}_t \right] \right\} + \\
  & & 
\;\;\;\;\;\;\; k_B \mbox{ln} Z_{t+\delta t}  - k_B \mbox{ln} Z_{t} - k_B \beta   \mbox{Tr}\left[ \rho(t ) \mathbfcal{F}  \right] \cdot \dot{\bf X}(t) \delta t. \nonumber \label{delst}
 \end{eqnarray}
 In the last Eq. we have used $\mbox{Tr}\left[ \rho(t+\delta t) \right] = \mbox{Tr}\left[ \rho(t) \right]=1$.
  We can identify the first term with a change in the internal energy,
  \begin{equation}
 U=\mbox{Tr}\left[ \rho(t) {\cal H}_t \right],
 \end{equation}
  i.~e. $\delta U=\mbox{Tr}\left[ \rho(t+\delta t) {\cal H}_t \right]-\mbox{Tr}\left[ \rho(t) {\cal H}_t \right]$, as well as the change in the internal free energy, 
  \begin{equation}
  F= - k_B T \; \mbox{ln} Z_{t}.
  \end{equation}
    The other terms are related to the work developed in the change of the time-dependent parameters \cite{balian},
 \begin{equation}
\delta W=  -\mbox{Tr}\left[ \rho(t ) \mathbfcal{F} \right] \cdot \dot{\bf X}(t) \delta t.
 \end{equation}
 Therefore, Eq. (\ref{delst}) can be  expressed as follows
 \begin{equation}
 T  \delta S[{\cal H}]=  \delta U  -  \delta F + \delta W.
 \end{equation}
 Following Ref. \cite{deff,kawai,horo,espo}, we define the non-equilibrium entropy production as the following difference
\begin{equation}
\delta S^{\rm neq} = \delta S - \delta S[{\cal H}],
 \end{equation}
 and we evaluate it for a protocol $\delta {\cal C}$  in the parameter space starting in ${\bf X}(t_0)$ and ending in ${\bf X}(\tau)$, which consists in a sequence of the previous small changes.
 Using Eq. (\ref{vn}) and (\ref{delst}), and introducing the definition of the relative entropy $S\left[\rho(t) ||\rho_t \right]= S(t)+k_B\mbox{Tr}\left[ \rho(t) \mbox{ln} \rho_{t} \right] $,
 the non-equilibrium entropy change can be written as in Ref. \cite{deff} 
 \begin{eqnarray}\label{sneq}
 \delta S^{\rm neq} &= &S\left[\rho({\tau}) ||\rho_{\tau} \right] - S\left[\rho(t_0) ||\rho_{t_0} \right]   \nonumber \\
 & & +k_B\beta \int_{\delta {\cal C}} \mathrm{d}t \mbox{Tr}\left[ \rho(t) \mathbfcal{F} \right] \cdot \dot{\bf X}(t).
 \end{eqnarray}

\section{Lehmann representation for the thermoadiabatic tensor}\label{lehmann}
Performing a Fourier transform in the adiabatic susceptibilities entering the of Eq. (\ref{lambda}), we see that the elements of this tensor can be expressed as
\begin{equation}\label{lambd}
\Lambda_{\mu,\nu}(\vec{ X}) = -i \partial_{\omega} \chi_{\mu,\nu}(\omega)|_{\omega=0}= \lim_{\omega \rightarrow 0} \frac{\mbox{Im}[\chi_{\mu,\nu}(\omega)]}{\omega},
\end{equation}
being $\chi_{\mu,\nu}(\omega)$ the Fourier transform of  the susceptibility $\chi_{\mu,\nu}(t-t^{\prime})=-i\theta(t-t^{\prime}) \langle \left[ {\cal F}_{\mu}(t), {\cal F}_{\nu}(t^{\prime})\right]\rangle_t$. 
Using  the notation ${\cal F}_{\mu} = - \partial_{\mu} {\cal H}_t$ and 
expressing the susceptibility in  the Lehmann representation we have
\begin{eqnarray} \label{chi}
\chi_{\mu,\nu}(\omega) &=&\hbar \sum_{n,m }  p_m \left( \varepsilon_m-\varepsilon_n\right)^2 
\left[ \frac{\langle \partial_{\mu} m|n\rangle \langle n| \partial_{\nu} m\rangle }{\omega - (\varepsilon_m-\varepsilon_n)+i \eta} \right. \nonumber\\
& & \left.  \;\;\;\;\;\;\;\;\;\;\;\;\;\;- \frac{\langle \partial_{\nu} m|n\rangle \langle n | \partial_{\mu} m\rangle }{\omega - (\varepsilon_n-\varepsilon_m)+i \eta} \right],
 \end{eqnarray}
 with  $\eta=0^+$. We have used the following identities calculated from ${\cal H}_t |n\rangle =\varepsilon_n |n\rangle$ and 
$\langle n| \partial_{\mu} \left( {\cal H}_t |m \rangle \right)$
\begin{eqnarray}\label{elem}
\langle n|\partial_{\mu} {\cal H}_t  |m\rangle & = & \left(\varepsilon_m -\varepsilon_n\right)\langle n|\partial_{\mu} m \rangle + \delta_{n,m} \partial_{\mu} \varepsilon_m, \nonumber \\
\langle m |\partial_{\mu} {\cal H}_t  |n \rangle & = & \left(\varepsilon_m -\varepsilon_n\right)\langle \partial_{\mu} m| n \rangle + \delta_{n,m} \partial_{\mu} \varepsilon_m,
\end{eqnarray}
Calculating the derivative as indicated in Eq. (\ref{lambd}), we have $\Lambda_{\mu,\nu}(\vec{ X})= \Lambda^A_{\mu,\nu}(\vec{ X})+\Lambda^S_{\mu,\nu}(\vec{ X})$, with
the antisymmetric and symmetric components given by
\begin{eqnarray}\label{lambf}
\Lambda^S_{\mu,\nu}(\vec{ X}) = 
& & \hbar \pi \lim_{\omega \rightarrow 0} \sum_{n,m}p_m  
\frac{(\varepsilon_n-\varepsilon_m)^2}{\omega}\mbox{Re}[\langle \partial_{\mu} m|n\rangle \langle n| \partial_{\nu} m\rangle ]\nonumber \\
& & \times \left[\delta(\omega-(\varepsilon_m-\varepsilon_n))- \delta(\omega-(\varepsilon_n-\varepsilon_m))\right]\nonumber \\
& & \Lambda^A_{\mu,\nu}(\vec{ X}) =   2 \hbar\sum_m p_m \;\mbox{Im} \left[\langle \partial_{\mu} m | \partial_{\nu} m \rangle\right],
\end{eqnarray}

\section{Driven qubit: Calculation of currents and power for different spin couplings}

\label{linearr}
\subsection{Coupling: $\hat{\tau}_L=\hat{\sigma}_x$ and $\hat{\tau}_R=\hat{\sigma}_z$}
The different components of $\mathbf{p}(t)$ for the driving protocol of Eq. (\ref{xb}) with $\hat{\tau}_L=\hat{\sigma}_x$ and $\hat{\tau}_R=\hat{\sigma}_z$ 
can be calculated by solving Eqs.~(\ref{insad1}) and (\ref{insad2}). They read
\begin{align}
\label{appprob}
&{p}_{1}^{\rm (i)}=\frac{1}{1+e^{-
\delta E/k_{\rm B}T}}, \nonumber \\
&{p}_{1,\Delta T}^{({\rm i})}=\frac{\delta E\Big(-{B_z^2}\Gamma_L+B_x^2\Gamma_R\Big)\sech^2\left(\frac{\delta E}{2k_{\rm B}T}\right)}{4k_{\rm B}\Big({B_z^2}\Gamma_L+B_x^2\Gamma_R\Big)} \frac{\Delta T}{T^2},\nonumber \\
&{p}_{1}^{({\rm a})}=-\frac{d{p}_1^{({\rm i})}}{dt}\frac{\delta E\tanh\left(\frac{\delta E}{2k_{\rm B}T}\right)e^{\delta E/\epsilon_{\rm C}}}{4\Big(B_z^2\Gamma_L+B_x^2\Gamma_R\Big)},\nonumber \\
&p_2^{\rm (i)}=1-p_{1}^{\rm (i)},\;\;{p}_{2,\Delta T}^{({\rm i})}=-{p}_{1,\Delta T}^{({\rm i})},\;\;{p}_{2}^{({\rm a})}=-{p}_{1}^{({\rm a})}.
\end{align}
The heat currents are
\begin{align}
&J_L^{({\rm a})}(t)=\frac{dp_1^{({\rm i})}}{dt}\frac{\delta E\, B_z^2\,\Gamma_L}{B_z^2\Gamma_L+B_x^2\Gamma_R},\nonumber \\
&J_R^{({\rm a})}(t)=\frac{dp_1^{({\rm i})}}{dt}\frac{\delta E\, B_x^2\,\Gamma_L}{B_z^2\Gamma_L+B_x^2\Gamma_R}.
\end{align}

\subsection{Coupling: $\hat{\tau}_L=\hat{\sigma}_x$ and $\hat{\tau}_R=\hat{\sigma}_y$}
In this case the adiabatic probabilities can be written as
\begin{align}
&p_{1}^{({\rm a})}=-\frac{dp_1^{({\rm i})}}{dt} \frac{\delta E e^{\delta E/\epsilon_{\rm C}}\tanh \left(\delta E/k_{\rm B}T\right)}{4B_z^2\Gamma_L+\delta E^2\Gamma_R},\nonumber \\
&p_{2}^{({\rm a})}=-p_1^{({\rm a})}.
\end{align}
In the absence of a bias, the instantaneous contribution to the current vanishes, and the only contributions come from adiabatic corrections. The adiabatic heat current flowing in the left and right lead are given by
\begin{align}
&J_{\rm L}^{({\rm a})}(t)=\frac{dp_1^{({\rm i})}}{dt} \frac{4\delta E B_{z}^2\,\Gamma_L}{4B_z^2\Gamma_L+\delta E^2 \Gamma_R},\nonumber \\
&J_{\rm R}^{({\rm a})}(t)=\frac{dp_1^{({\rm i})}}{dt} \frac{\delta E^3\Gamma_R}{4B_z^2\Gamma_L+\delta E^2\Gamma_R}.
\label{curradiabatic}
\end{align}
Using the modulation in Eq.~(\ref{xb}) with $\phi =\frac{\pi}{2}$, we obtain
\begin{align}
&\frac{dp_1^{({\rm i})}}{dt}=\frac{-\Omega \sech^2\left(\delta E/k_{\rm B T}\right)}{2k_{\rm B}T\delta E} \Big[2B_{z,0} B_{z,1} \sin(\Omega t)\nonumber\\
&~~~~~~~~~+2B_{x,0}B_{x,1}\cos(\Omega t) +\Big(B_{z,1}^2-B_{x,1}^2\Big)\sin(2\Omega t)\Big].
\label{dpt}
\end{align}
Plugging Eq.~(\ref{dpt}) into Eqs.~(\ref{curradiabatic}), the time averaged adiabatic heat currents can be written as a function of different parameters
\begin{multline}
J_{\rm tr,L}^{{\rm Q}}=\frac{1}{\tau}\int_0^\tau J_{\rm L}^{({\rm a})} (t)\\
=\frac{k_{\rm B}T\Omega}{2\pi}\int_0^{2\pi}dx f\Big[\frac{\epsilon_0}{k_{\rm B}T},\frac{\epsilon_1}{k_{\rm B}T},\frac{\Delta_0}{k_{\rm B}T},\frac{\Delta_1}{k_{\rm B}T},\frac{\Gamma_\alpha}{k_{\rm B}T},x\Big],
\label{linear}
\end{multline} 
where $f$ is a dimensionless function which depends on all the parameters of the driving modulation and on the coupling strengths with the leads. Similar expression can be obtained for the heat current flowing in the right contact. In particular, the adiabatic heat currents are linear in the driving frequency as observed in Eq.~(\ref{linear}).

\subsection{Symmetry properties of $\Lambda_{\ell,\ell^{\prime}}$}
For $\Delta T =0$, we can rewrite the work $W$ as
\begin{equation}
W=\int_0^{2\pi/\Omega} dt \Big[\frac{dE_1}{dt}p_1^{({\rm a})}+\frac{dE_2}{dt}p_2^{({\rm a})}\Big]
\end{equation}
and, by using the normalization condition $\sum_jp_j^{({\rm a})}=0$ and the fact that $E_1(t)=-E_2(t)$, we find
\begin{equation}
W=2\sum_{j}\int_0^{2\pi/\Omega} dt \;\frac{dE_2}{dX_j}\dot{X}_jp_{2}^{(a)},
\label{poweraa}
\end{equation}
where $X_1(t)$ and $X_2(t)$ are the two driving parameters of the q-bit. Moreover, applying the fact that $\delta E=2E_2$, we find
\begin{align}
W&=\int_0^{2\pi/\Omega} dt \;{\zeta}(\mathbf{B})\sum_{j,k}\frac{d\delta E}{dX_j}\frac{dp_2^{({\rm i})}}{dX_k}\dot{X}_{j}\dot{X}_k \nonumber \\
&=\int_0^{2\pi/\Omega} dt \;{\zeta}(\mathbf{B})\sum_{j,k}\frac{d\delta E}{dX_j}\frac{dp_2^{({\rm i})}}{d\delta E}\frac{d\delta E}{dX_k}\dot{X}_{j}\dot{X}_k
\label{poweraa}
\end{align}
where ${\zeta}(\mathbf{B})$ is defined by the relation $p_2^{({\rm a})}=\zeta(\mathbf{B})\frac{dp_2^{({\rm i})}}{dt}$ (see Eq.~(\ref{appprob})), which is a consequence of Eq.~(\ref{papi}). Comparing Eq.~(\ref{poweraa}) with Eq.~(\ref{worklambda}), we obtain:
\begin{equation}\label{aplamqbit}
\Lambda_{12}(\mathbf{B})=\Lambda_{21}(\mathbf{B})={\zeta}(\mathbf{B})\frac{d\delta E}{dX_1}\frac{dp_{2}^{({\rm i})}}{d\delta E}\frac{d\delta E}{dX_2}.
\end{equation}
Eq.~(\ref{aplamqbit}) and Eq.~(\ref{lambdallp}) have the same form.

\section{Driven quantum dot - calculation of the thermal geometric tensor}\label{aplam}
We need to calculate the following coefficients
\begin{eqnarray}\label{lambdalj}
\Lambda_{\mu,\nu}(t) &=& \frac{1}{\hbar}  \int_{-\infty}^{-\infty} d t^{\prime} (t-t^{\prime}) \chi_{\mu,\nu}(t-t^{\prime}) \nonumber \\
&=& -\lim_{\omega \rightarrow 0} \frac{\mbox{Im}\left[ \chi_{\mu,\nu}(\omega) \right] }{\hbar \omega}
,\;\;\mu,\nu=1,2,3
\end{eqnarray}
with
\begin{equation}
\chi_{\mu,\nu}(t-t^{\prime})= - i \theta(t-t^{\prime}) \langle \left[ {\cal F}_\mu(t), {\cal F}_\nu(t^{\prime}) \right] \rangle,
\end{equation}
being $ {\cal F}_{1,2}=\Psi^{\dagger}_d \; \hat{\sigma}_{x,z} \Psi_d$, and ${\cal F}_3={\cal J}_{Q,R}= -i \sum_{k_R,s,\sigma}\varepsilon_{k_R,s} v_{k_R,s,\sigma}
c^{\dagger}_{k_R,s} d_{\sigma} + H. c. $. We can calculate (\ref{lambdalj}) following standard procedures based on the formalism of  imaginary-time Green's functions. 
   We can define
$\hat{\cal G}(\tau) = - \langle T_{\tau} \left[ \Psi_d(\tau) \Psi^{\dagger}_d(0) \right] \rangle$ and 
$\hat{\cal G}_{k_R,d}(\tau) = - \langle T_{\tau} \left[ \Psi_{k_R} (\tau) \Psi^{\dagger}_d(0) \right] \rangle$, where $ T_{\tau} $ denotes ordering along the imaginary axis. 
In terms of this, it is possible to write
\begin{eqnarray}\label{chimat}
\chi_{\ell,\ell^{\prime}}(iq_n) &=&\frac{1}{\beta} \sum_{ik_n} \mbox{Tr} \left[\hat{\sigma}_\ell \hat{\cal G}(ik_n+iq_n) \hat{\sigma}_{\ell^{\prime}} \hat{\cal G}(ik_n)\right],\nonumber \\
\chi_{3,\ell}(iq_n) &=&\frac{1}{\beta} \sum_{ik_n}\sum_{k_R} \mbox{Tr} \left\{  \hat{\varepsilon}_{k_R} \hat{v}_R \left[ i \hat{\cal G}(ik_n+iq_n) \hat{\sigma}_\ell 
\right. \right.
\nonumber \\
& & \left. \left. \times \hat{\cal G}_{d,k_R}(ik_n) -i\hat{\cal G}_{k_R,d}(ik_n) \hat{\sigma}_\ell \hat{\cal G}(ik_n-iq_n) \right] \right\} \nonumber \\ 
&=& - \chi_{\ell,3}(iq_n), \;\;\;\;\;\; \;\ell,\ell^{\prime}=1,2,\nonumber \\
\chi_{3,3}(iq_n) &=&- \frac{1}{\beta} \sum_{ik_n}\sum_{k_R, k^{\prime}_L} \nonumber \\
& & \mbox{Tr} \left\{  \hat{\varepsilon}_{k_R} \hat{v}_R 
  \hat{\cal G}_{k_R,d}(ik_n+iq_n)\hat{\varepsilon}_{k^{\prime}_R} \hat{v}_L \hat{\cal G}_{k^{\prime}_L,d}(ik_n) 
\right. 
\nonumber \\
& & \left. 
 + \hat{\cal G}_{d,k_R}(ik_n+iq_n)\hat{\varepsilon}_{k_R} \hat{v}_R 
  \hat{\cal G}_{d,k^{\prime}_L}(ik_n) 
  \hat{\varepsilon}_{k^{\prime}_L}  \hat{v}_L
  \right\} \end{eqnarray}
with $\varepsilon_{k_{\alpha},s,s^{\prime}}= \varepsilon_{k_{\alpha},s} \delta_{s,s^{\prime}}$, $\alpha=L,R$, $q_n= 2 \pi n/\beta$ and $k_n= (2n+1) \pi /\beta$. 

It is convenient to introduce the spectral representation
\begin{eqnarray}\label{greenmat}
& & \hat{\cal G}(ik_n)= \int \frac{ d\varepsilon}{2 \pi} \frac{\hat{\rho}_t(\varepsilon)}{ik_n- \varepsilon}, \nonumber \\
& &\hat{\cal G}_{k_{\alpha},d}(ik_n)= \int \frac{ d\varepsilon}{2 \pi} \frac{\hat{\rho}_{k_{\alpha},d}(\varepsilon)}{ik_n- \varepsilon}
\end{eqnarray}
with
\begin{eqnarray}
& & \hat{\rho}_t(\varepsilon)= -2 \mbox{Im}\left[\hat{G}_t(\varepsilon)\right]=
\hat{G}_t(\varepsilon)\hat{\Gamma} \left[\hat{G}_t(\varepsilon)\right]^{\dagger},  \\
& & \hat{\rho}_{k_{\alpha},d}(\varepsilon)=  \hat{\rho}^0_{k_{\alpha}}(\varepsilon) \hat{v}_{\alpha} \hat{\rho}_t(\varepsilon)+ \hat{\rho}^0_{k_{\alpha}}(\varepsilon) \hat{v}_{\alpha} \hat{\rho}_t (\varepsilon),
\end{eqnarray}
where $\rho^0_{k_{\alpha}, s,s^{\prime}}(\varepsilon)= 2 \pi \delta_{s,s^{\prime}} \delta(\varepsilon-\varepsilon_{k_{\alpha},s})$ and
$G^0_{k_{\alpha},s,s^{\prime}}(\varepsilon) = \delta_{s,s^{\prime}} \left( \varepsilon + i \eta-\varepsilon_{k_{\alpha},s} \right)^{-1}$.
The retarded frozen Green's function of the quantum dot in contact to the reservoirs is given in Eq. (\ref{gdot}),
 while $\hat{\Gamma}= - 2 \mbox{Im}\left[\hat{G}_t(\varepsilon)^{-1}\right]=
\sum_{\alpha}\hat{\Gamma}_{\alpha}$
is the hybridization matrix accounting for the contact between the quantum dot and the reservoirs, being $ \hat{\Gamma}_{\alpha} = \sum_{k_{\alpha}} \hat{v}_{k_{\alpha}} \hat{\rho}^0_{k_{\alpha}}\hat{v}_{k_{\alpha}} $.

Using  Eq. (\ref{greenmat}) into Eq. (\ref{chimat}), after some algebra and performing the analytic continuation to the real axis we get
\begin{eqnarray}
\Lambda_{\ell,{\ell^{\prime}}}(t) &=& - \frac{1}{h} \int d \varepsilon \frac{df (\varepsilon)}{d\varepsilon} \mbox{Tr} \left[\hat{\sigma}_{\ell} \hat{\rho}_t(\varepsilon) \hat{\sigma}_{\ell^{\prime}} \hat{\rho}_t(\varepsilon) \right],\;\ell,\ell^{\prime}=1,2 \nonumber \\
\Lambda_{3,\ell}(t) &=& - \frac{1}{h} \int d \varepsilon \; \varepsilon \frac{df (\varepsilon)}{d\varepsilon} \mbox{Tr} \left[\hat{\Gamma}_R \hat{\rho}_t(\varepsilon) \hat{\sigma}_\ell \hat{\rho}_t(\varepsilon) \right],\nonumber \\
&=& -  \Lambda_{\ell,3}(t) \;\;\;\;\;\; \ell=1,2, \nonumber \\
\Lambda_{3,3}(t) &=& - \frac{1}{h} \int d \varepsilon \; \varepsilon^2 \frac{df (\varepsilon)}{d\varepsilon} \mbox{Tr} \left[\hat{\Gamma}_R \hat{G}_t(\varepsilon) \hat{\Gamma}_L \hat{G}_t^{\dagger}(\varepsilon)\right],
\end{eqnarray}

\bibliography{adiabatic_geo_resub}

\end{document}